\documentclass[draft, jgrga]{agu2001}
\usepackage{graphicx}
\usepackage{amssymb,amsmath}
\usepackage{color}

\authorrunninghead{CAMPOREALE et al.}

\titlerunninghead{Lower hybrid to whistler mode conversion}

\authoraddr{E. Camporeale: enrico@lanl.gov}

\begin{document}

%
%
%
\setkeys{Gin}{draft=false}
%
%

%
%

\title{Lower hybrid to whistler mode conversion on a density striation}

%
%
\author{E. Camporeale, G.~L. Delzanno, P. Colestock}
\affil{Los Alamos National Laboratory, 87545 Los Alamos, NM, USA.}

%
%


\begin{abstract}
When a wave packet composed of short wavelength lower hybrid modes traveling in an homogeneous plasma region encounters an inhomogeneity, 
it can resonantly excite long wavelength whistler waves via a linear mechanism known as mode conversion. An enhancement of lower hybrid/whistler activity has been often observed by sounding rockets and satellites in the presence of 
density depletions (striations) in the upper ionosphere. We address here the process of linear mode conversion of lower hybrid 
to whistler waves, mediated by a density striation, using a scalar-field formalism (in the limit of cold plasma linear theory) which we solve numerically.   
We show that the mode conversion can effectively transfer a large amount of energy from the 
short to the long wavelength modes.   We also study how the efficiency scales by changing the properties (width and amplitude) of the density striation. 
 We present a general criterion for the width of the striation that, if fulfilled, maximizes the conversion efficiency.  
Such a criterion could provide an interpretation of recent laboratory experiments carried out on the Large Plasma Device at UCLA.
\end{abstract}

%
%

%

\begin{article}

%
%

\section{Introduction}\label{intro}

Whistler and lower hybrid waves are frequently observed in all regions of the Earth's magnetosphere. 
They belong to the same branch of the dispersion relation for waves in a magnetized plasma.  
Whistler modes are electromagnetic waves with frequency between the ion and electron cyclotron frequencies,
 whereas lower hybrid waves are predominantly electrostatic and their frequency approaches the lower hybrid frequency in the limit of infinite wavevector for exactly perpendicular propagation.  
The close relationship between these two wave modes suggests that linear mode conversion can likely occur wherever inhomogeneities are present.

In the ionosphere, enhancement of lower hybrid/whistler wave activity  has been often associated with density depletions, so-called lower hybrid cavities (LHCs) or lower hybrid solitary wave structures (LHSS).
Both sounding rockets \citep{labelle86, delory97, mcadams98, schuck98} 
and satellites \citep{eriksson94, ergun95, hoymork00, tjulin03, tjulin04, reiniusson06} have observed intense, 
primarily electrostatic waves in the presence of density cavities. The depth of the density depletions measured in the rocket experiments were as large as 80\%, 
although measurements by the Freja satellite \citep{eriksson94, ergun95, reiniusson06} suggested that the depletions were much shallower, typically a few per cent.
  Lower hybrid cavities were also observed by the Viking and Cluster satellites at much higher altitudes \citep{tjulin03, tjulin04}, with depletions ranging from a few to 30\% of the background density.   
A statistical survey of Freja measurements by \citet{hoymork00} has shown that in a vast majority of cases the density depletion is well fitted by a Gaussian shape 
and the average cavity width was on the order of 30 meters, corresponding to few ion gyroradii. Theoretical studies showed how lower hybrid waves could be trapped inside such structures and differences 
in the observed features of the wave modes above and below the lower hybrid frequency could be explained \citep{seyler94, schuck98, borisov08}.  In addition,  nonlinear theory 
indicates that the cavities could collapse due to the ponderomotive force of the lower hybrid waves trapped inside [e.g., \citet{shapiro93}].\\
While a complete theory that would explain the formation of the cavities and the propagation of waves inside them is still missing, some observations showing circularly polarized low frequency waves 
support the idea that the density gradients are preformed and that whistler waves impinging on a density cavity or gradient can mode convert to lower hybrid waves \citep{delory97, reiniusson06}. 
 This process was originally studied by \citet{bell90} to explain observations of electrostatic waves excited by whistlers throughout the ionosphere and magnetosphere, and this work was recently augmented in \citet{foust10}. 
Formal solutions of the wave equations in a cold magnetized plasma showed that the process involves a transmitted and reflected whistler wave and two lower hybrid waves.  
In addition, linear mode conversion implies that the inverse process is also possible: an incident LH wave can mode convert to a whistler mode.  \citet{borisov95} proposed 
this mechanism in ionospheric modification experiments to explain the conversion of lower hybrid waves that are produced by the decay of HF radio waves  to whistler waves on density irregularities.   
He estimated a very small conversion efficiency ($\approx 10^{-6}$) for typical experimental parameters.\\
In addition to observations, several laboratory experiments have addressed the physics of linear mode conversion, mimicking the conditions found in the magnetosphere.  Experiments at the Large Plasma Device (LAPD) at UCLA 
have investigated the conversion of whistler to lower hybrid waves \citep{bamber94,bamber95}, and of lower hybrid to whistler modes \citep{rosenberg98, rosenberg00, rosenberg01,vanCompernolle11} in a cylindrical field-aligned density depletion. 
None of these works, however, have focused on the conversion efficiency, which is the primary goal of this paper.\\
Recently, \citet{eliasson08} have presented a numerical study of the mode conversion between lower hybrid and whistler waves on one or more density striations. 
Their theory is valid in the cold plasma limit and for shallow striations, and they have presented an empirical criterion for the width of the striation that, if satisfied, would maximize the efficiency of the conversion.
The same formalism and numerical method has been used to study the problem of whistler wave attenuation on density striations (the so-called '20dB puzzle') by \citet{shao12}.\\
In this paper, we address the problem of linear mode conversion between lower hybrid and whistler modes on a density striation from a theoretical standpoint. 
The problem is set up as the following. A wave-packet composed of quasi-perpendicular lower hybrid modes is 
initialized in a region of uniform magnetized plasma. The packet propagates according to its group velocity until it encounters a depletion in the density, in the direction perpendicular to the
background magnetic field. The effect of the density striation is to couple linear modes that would otherwise be uncoupled in the absence of any inhomogeneities. This is a completely linear mechanism, that does not need any non-linear effect. 
Hence, a fraction of the energy of the lower hybrid
packet is transferred to whistler waves. The scope of the present paper is to analyze what is the efficiency of such energy transfer, what are the variables that affect
such efficiency, and under which conditions the efficiency is maximum. \\
The results presented here are complementary to those of \citet{eliasson08}. Both approaches start from the linearized cold plasma approximation, although we use a formulation in terms of scalar fields. 
However, our approach is valid for striations of arbitrary depth and we study density striations of up to 80\% of the background. This regime is relevant to recent experiments performed on LAPD \citep{vanCompernolle11}.
More importantly, we study mode conversion in a wider parameter space and we will derive a general criterion for the width of the striation that includes the one proposed by \citet{eliasson08} as a special case.\\
The paper is organized as follows. We describe the mathematical model in Section 2. The numerical results and the discussion on the conversion efficiency  and on the scaling laws obtained by varying some of the system parameters are presented in Section 3. The conclusions and future research directions are drawn in Section 4.\\

\section{Mathematical model}\label{model}
We consider a plasma composed of electrons and singly charged ions, in a Cartesian geometry $(x,\,y,\,z)$ where the $z$ direction is ignorable, namely $\partial /\partial z=0$.
The plasma is described by the continuity and momentum equations, and by Maxwell's equations. 
Those equations are linearized relative to an equilibrium characterized by an inhomogeneous plasma density, $n_{e,\,eq}=n_{i,\,eq}=n_{eq}(x,y)$, 
zero plasma flow, ${\bf V}_{e,\,eq}={\bf V}_{i,\,eq}=0$, and a homogeneous magnetic field, 
${\bf B}_{eq}=(B_{x,\,eq},\,B_{y,\,eq},\,0)$ with no guide
field. Subscripts $e$ and $i$ label electrons and ions respectively. 
The model equations therefore are
\begin{eqnarray}
&& \frac{\partial n}{\partial t}+\nabla \cdot \left( n_{eq} {\bf V}_i\right)=0,\label{cont} \\
&& m_i \frac{\partial {\bf V}_i}{\partial t}=e {\bf E} + e {\bf V}_i \times {\bf B}_{eq},\label{momi}\\
&& m_e \frac{\partial {\bf V}_e}{\partial t}=-e {\bf E} - e {\bf V}_e \times {\bf B}_{eq},\label{mome}\\
&& \frac{\partial {\bf B}}{\partial t}=-\nabla \times {\bf E},\label{far}\\
&& \nabla \times {\bf B}=\mu_0 e n_{eq}\left({\bf V}_i-{\bf V}_e\right),\label{amp}\\
&& \nabla \cdot {\bf B}=0, \label{div}
\end{eqnarray}
where $n$ is the perturbed plasma density, ${\bf V}_{e,\,i}$ are the perturbed electron and ion flows, 
${\bf E}$ is the perturbed electric field and ${\bf B}$ is the perturbed magnetic field.
In Eqs. (\ref{cont})-(\ref{div}), $e$ is the positive elementary charged, $m_{e,\,i}$ are the electron and ion masses,
and $\mu_0$ is the permeability of vacuum. Equations (\ref{cont})-(\ref{div}) are derived under the following
assumptions: 
\begin{enumerate}[i)]
\item the frequency associated with perturbed quantities are much smaller than the electron plasma frequency. Therefore, perturbations are quasi-neutral, namely $n_e=n_i=n$, and hence Poisson's equation
is not included in our model;
\item the plasma is cold, namely pressure terms are dropped in Eqs. (\ref{momi})
and (\ref{mome});
\item consistent with i), the displacement current is neglected in Ampere's law (\ref{amp}). 
\end{enumerate}
We notice that in the region of interest (upper ionosphere and plasmasphere), the cold plasma approximation is justified since the value of the plasma beta is very small (around $10^{-5}$) \citep{kelley_book}.
We neglect collisions between ions and electrons and between charged particles and neutrals, therefore we do not include an equation for neutrals. This assumption is again justified when using 
the model for the upper ionosphere, or the plasmasphere (see Section 3).
Following \citet{delzanno04}, we introduce the following representation of the vector fields
in terms of scalar fields
\begin{eqnarray}
&& {\bf V}_i=\nabla \times \left(\varphi{\bf e}_z\right)+\nabla \chi + V_z {\bf e}_z \label{vi} \\
&& {\bf B}=\nabla \times \left(\psi{\bf e}_z\right)+ B_z {\bf e}_z \label{b}
\end{eqnarray}
where ${\bf e}_z$ is the unit vector along $z$. 
We note that Eq. (\ref{b}) implies that Eq. (\ref{div}) is automatically satisfied for our system with
$\partial /\partial z=0$. In addition, by employing a vector potential formulation for the electromagnetic
field, ${\bf B}=\nabla \times {\bf A}$ and ${\bf E}=-\nabla \phi -
\frac{\partial {\bf A}}{\partial t}$ (with ${\bf A}$ and $\phi$ the perturbed vector
and electrostatic potentials, respectively), which guarantees that Faraday's law (\ref{far}) is satisfied,
one can see that
\begin{equation}
A_z=\psi \label{az}
\end{equation}
and
\begin{equation}
E_z=-\frac{\partial \psi}{\partial t}. \label{ez}
\end{equation}
We also note that $\nabla \cdot {\bf V}_i=\nabla^2 \chi$, implying that compressibility
is associated with the $\chi$ field. 

We introduce the following normalization:
\begin{equation}
t\longrightarrow \omega_{lh}t,\,{\bf x} \longrightarrow \frac{{\bf x}}{d_e},\,
n \longrightarrow \frac{n}{n_0},\,
{\bf B} \longrightarrow \frac{{\bf B}}{B_0},\,{\bf E} \longrightarrow \frac{{\bf E}}{\omega_{lh}d_e B_0}.
\end{equation}
Consistently,
\begin{equation}
{\bf V} \longrightarrow \frac{{\bf V}}{\omega_{lh} d_e},\,
\varphi\longrightarrow \frac{\varphi}{\omega_{lh}d_e^2},\,
\chi\longrightarrow \frac{\chi}{\omega_{lh}d_e^2},\,
\psi \longrightarrow \frac{\psi}{d_e B_0}.
\end{equation}
Here, $n_0$ and $B_0$ are some reference density and magnetic field, $\omega_{lh}$ is the lower hybrid
frequency in the limit $\omega_{pe}\gg\omega_{ce}$($\omega_{lh}=\sqrt{\omega_{ce} \omega_{ci}}$, with $\omega_{ce}=e B_0/m_e$ the electron cyclotron
frequency and similarly for the ions) and $d_e=c/\omega_{pe}$ is the electron inertial length ($c$ is the
speed of light and $\omega_{pe}$ is the electron plasma frequency).

Algebraic manipulations of Eqs. (\ref{cont})-(\ref{div}) lead to
a closed system of equations for the five scalar fields $\varphi,\,V_z,\,\psi,\,B_z,\,\chi$. 
Specifically, by applying ${\bf e}_z \cdot \nabla \times$
to Eq. (\ref{momi}) one obtains
\begin{equation}
\frac{\partial }{\partial t} \nabla^2 \varphi-\frac{\omega_{ci}}{\omega_{lh}}
\frac{\partial B_z}{\partial t}=-\frac{\omega_{ci}}{\omega_{lh}} {\bf B}_{eq} \cdot \nabla V_z.
\label{fi}
\end{equation}
The $z$ component of Eq. (\ref{momi}) leads to
\begin{equation}
\frac{\partial V_z}{\partial t}+\frac{\omega_{ci}}{\omega_{lh}} \frac{\partial \psi}{\partial t}=
\frac{\omega_{ci}}{\omega_{lh}}\left({\bf B}_{eq} \cdot \nabla \varphi +
{\bf e}_z \cdot \nabla \chi \times {\bf B}_{eq}\right).
\label{vz}
\end{equation}
On the other hand, ${\bf e}_z \cdot$ applied to Eq. (\ref{mome}) leads to 
\begin{equation}
\frac{\partial \psi}{\partial t}-\frac{1}{n_{eq}}\frac{\partial}{\partial t} \nabla^2 \psi=
-\frac{\omega_{ce}}{\omega_{lh}}\frac{1}{n_{eq}}{\bf B}_{eq}\cdot \nabla B_z+{\bf B}_{eq}\cdot \nabla \varphi
+{\bf e}_z \cdot \nabla \chi \times {\bf B}_{eq}.
\label{psi}
\end{equation}
In deriving Eq. (\ref{psi}), we have expressed ${\bf V}_e$ in terms of the scalar fields through Ampere's
law (\ref{amp}).
Similarly, ${\bf e}_z \cdot \nabla \times$ applied to Eq. (\ref{mome}) leads to
\begin{equation}
\frac{\partial B_z}{\partial t}-\frac{1}{n_{eq}}\frac{\partial}{\partial t} \nabla^2 B_z+
\frac{1}{n_{eq}^2}\nabla n_{eq} \cdot \nabla \frac{\partial B_z}{\partial t}=
{\bf B}_{eq} \cdot \nabla V_z+\frac{\omega_{ce}}{\omega_{lh}}\frac{1}{n_{eq}}{\bf B}_{eq}\cdot 
\nabla \nabla^2 \psi-\frac{\omega_{ce}}{\omega_{lh}}\frac{1}{n_{eq}^2}{\bf B}_{eq} \cdot \nabla n_{eq} 
\nabla^2 \psi.
\label{bz}
\end{equation}
Finally, by adding Eqs. (\ref{momi}) and (\ref{mome}) and applying $\nabla \cdot$ to the resulting
equation, one obtains
\begin{equation}
\frac{\partial}{\partial t}\nabla^2 \chi+\frac{\omega_{ci}}{\omega_{lh}}\frac{1}{n_{eq}^2}
{\bf e}_z \cdot \nabla n_{eq} \times \nabla \frac{\partial B_z}{\partial t}=
-\frac{1}{n_{eq}}{\bf e}_z \cdot {\bf B}_{eq} \times \nabla \nabla^2 \psi
-\frac{1}{n_{eq}^2}{\bf e}_z \cdot \nabla n_{eq} \times {\bf B}_{eq} \nabla^2 \psi.\label{chi}
\end{equation}
Note that in the derivation of Eqs. (\ref{psi})-(\ref{chi}) we have used $1+m_e/m_i\simeq 1$.
Equations (\ref{fi})-(\ref{chi}) are sufficient to study the linear dynamics of the system. Other quantities
such as $n$ or $\phi$ can be obtained by post-processing once the solution for Eqs. (\ref{fi})-(\ref{chi}) is known. We also note that the effect of
equilibrium plasma density gradients enters only through Eqs. (\ref{psi})-(\ref{chi}).

Following \citet{eliasson08}, we specialize Eqs. (\ref{fi})-(\ref{chi}) to the case ${\bf B}_{eq}=(1,\,0,\,0)$ and with equilibrium density gradient
perpendicular to the equilibrium magnetic field, $\nabla n_{eq}=n^\prime_{eq}(y) {\bf e}_y$ (where prime means derivative with respect to $y$
and ${\bf e}_y$ is the unit vector along $y$). It follows that
\begin{eqnarray}
&& \frac{\partial }{\partial t} \nabla^2 \varphi-\frac{\omega_{ci}}{\omega_{lh}}
\frac{\partial B_z}{\partial t}=-\frac{\omega_{ci}}{\omega_{lh}} \frac{\partial V_z}{\partial x}, \label{fi2} \\
&& \frac{\partial V_z}{\partial t}+\frac{\omega_{ci}}{\omega_{lh}} \frac{\partial \psi}{\partial t}=
\frac{\omega_{ci}}{\omega_{lh}}\left(\frac{\partial \varphi}{\partial x}
-\frac{\partial \chi}{\partial y}\right), \label{vz2} \\
&& n_{eq}\frac{\partial \psi}{\partial t}-\frac{\partial}{\partial t} \nabla^2 \psi=
-\frac{\omega_{ce}}{\omega_{lh}}\frac{\partial B_z}{\partial x} +n_{eq}\frac{\partial \varphi}{\partial x}
-n_{eq}\frac{\partial \chi}{\partial y}, \label{psi2} \\
&& n_{eq}^2\frac{\partial B_z}{\partial t}-n_{eq}\frac{\partial}{\partial t} \nabla^2 B_z+
n^\prime_{eq}(y)\frac{\partial^2 B_z}{\partial t \partial y}=n_{eq}^2
\frac{\partial V_z}{\partial x}+\frac{\omega_{ce}}{\omega_{lh}}n_{eq}
\frac{\partial \nabla^2 \psi}{\partial x}, \label{bz2} \\
&&n_{eq}^2\frac{\partial}{\partial t}\nabla^2 \chi-\frac{\omega_{ci}}{\omega_{lh}}n^\prime_{eq}(y)
\frac{\partial^2 B_z}{\partial t \partial x}=
-n_{eq}\frac{\partial \nabla^2 \psi}{\partial y}
+n^\prime_{eq}(y)\nabla^2 \psi. \label{chi2}
\end{eqnarray}
Equations (\ref{fi2})-(\ref{chi2}) are the focus of our analysis in Sec. \ref{cp}.

\subsection{Waves in a homogeneous plasma}
Before analyzing the effect of density inhomogeneities (i.e. density striations) on wave propagation, we briefly discuss
the case of a homogeneous plasma and identify the properties of lower hybrid and whistler waves in terms of the
scalar fields introduced above. We set $n_{eq}=1$ ($\nabla n_{eq}=0$) and consider perturbations of the form $h =\hat{h} \exp
\left(-i \omega t + i k_\parallel x + i k_\perp y\right)$ where $h$ is a generic unknown, $\omega$ is the eigenfrequency and $k_\parallel$ ($k_\perp$) is
the wavenumber along $x$ ($y$). Equations (\ref{fi})-(\ref{chi}) become
\begin{eqnarray}
&& \omega \left(k_\parallel^2 + k_\perp^2\right) \hat{\varphi}+\omega \frac{\omega_{ci}}{\omega_{lh}}\hat{B}_z
= -\frac{\omega_{ci}}{\omega_{lh}} k_\parallel \hat{V}_z, \label{fip} \\
&& \omega \hat{V}_z+\omega \frac{\omega_{ci}}{\omega_{lh}} \hat{\psi}=
-\frac{\omega_{ci}}{\omega_{lh}}\left(k_\parallel \hat{\varphi}-k_\perp \hat{\chi} \right) \label{vzp}, \\
&& \omega \left(1+k_\parallel^2+k_\perp^2 \right) \hat{\psi}=\frac{\omega_{ce}}{\omega_{lh}} k_\parallel \hat{B}_z
-k_\parallel\hat{\varphi} + k_\perp \hat{\chi} \label{psip}, \\
&& \omega \left(1+k_\parallel^2+k_\perp^2 \right) \hat{B}_z=-k_\parallel \hat{V}_z+
\frac{\omega_{ce}}{\omega_{lh}} k_\parallel \left(k_\parallel^2+k_\perp^2 \right) \hat{\psi}, \label{bzp} \\
&& \omega \hat{\chi}=k_\perp \hat{\psi}. \label{chip}
\end{eqnarray}

For perpendicular propagation, $k_\parallel=0$ and $k_\perp\ne0$, it is easy to recover lower hybrid waves,
\begin{equation}
\omega^2=\frac{k_\perp^2}{1+k_\perp^2},
\label{olh}
\end{equation}
with $\omega \longrightarrow 1$ as $k_\perp \longrightarrow \infty$ (recall that $\omega$ is normalized to
the lower hybrid frequency and $k_\perp$ to the electron inertial length). Furthermore, 
\begin{equation}
\hat{\chi}=\sqrt{1+k_\perp^2}\hat{\psi}, 
\end{equation}
showing that the mode becomes electrostatic as
$k_\perp \gg 1$, and 
\begin{equation}
\hat{V}_z=\frac{\omega_{ci}}{\omega_{lh}}k_\perp^2\hat{\psi} \label{vzlh}
\end{equation}
while $\hat{\varphi}=\hat{B}_z=0$.

For parallel propagation, $k_\parallel\ne0$ and $k_\perp=0$, one can recover whistler waves: in the limit 
$\frac{\omega_{ci}}{\omega_{lh}} \ll \omega \ll \frac{\omega_{ce}}{\omega_{lh}}$, one obtains
\begin{equation}
\omega\simeq\frac{\omega_{ce}}{\omega_{lh}}k_\parallel^2
\label{whistler}
\end{equation}
in addition to the following relations between the various quantities
\begin{eqnarray}
&& \hat{\varphi}\simeq -\frac{\omega_{ci}}{\omega_{lh}}\frac{1}{k_\parallel} \hat{\psi}\label{fi3} \\
&& \hat{V}_z\simeq -\frac{\omega_{ci}}{\omega_{lh}} \hat{\psi} \label{vz3} \\
&& \hat{B}_z\simeq k_\parallel \hat{\psi} \label{bz3}
\end{eqnarray}
and $\hat{\chi}=0$.
\subsection{Mode Conversion}
\begin{figure}
\noindent\includegraphics[width=20pc]{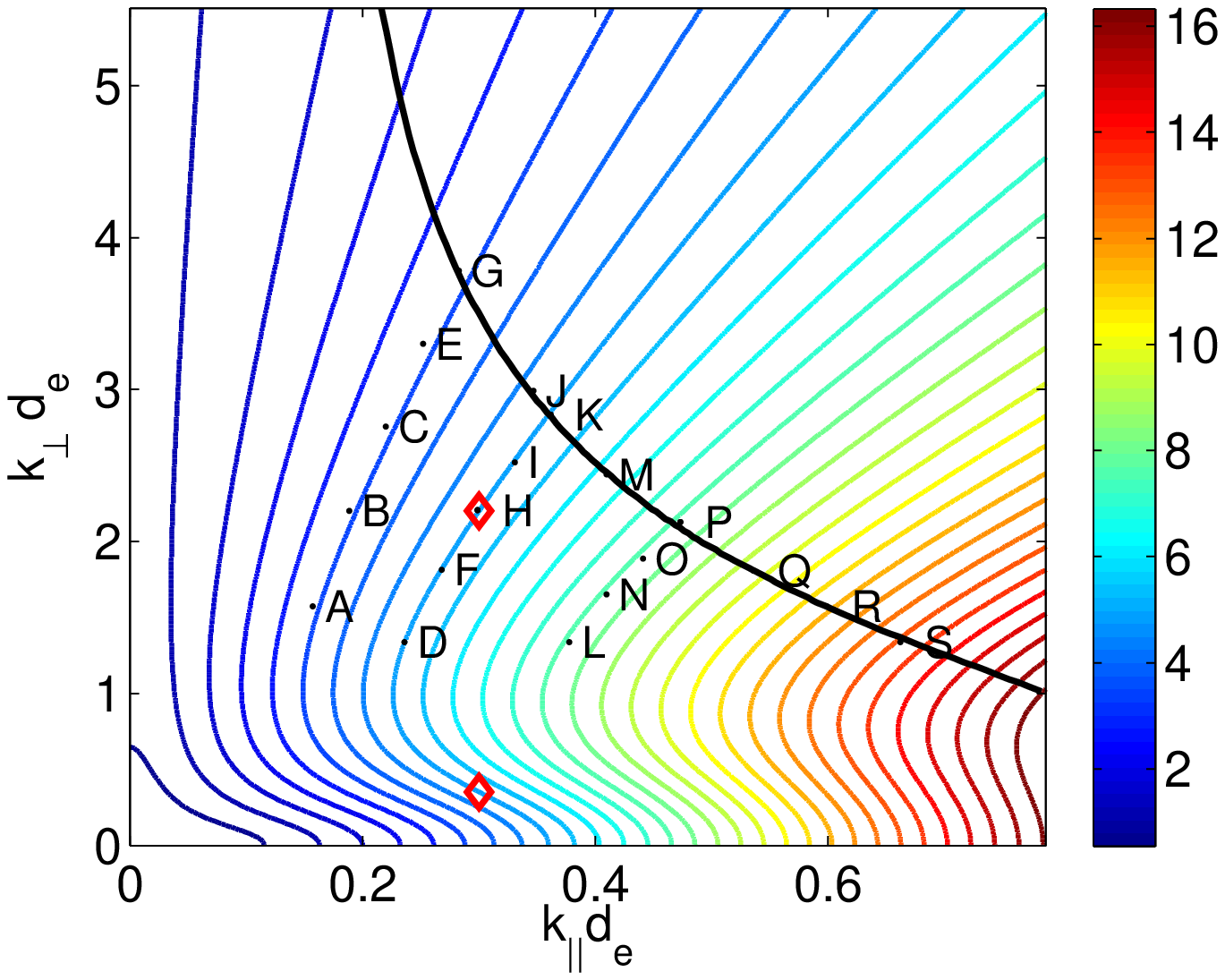}
 \caption{Contour plot of the frequency $\omega$ in $(k_\parallel,k_\perp)$ space from the dispersion relation obtained by Eqs.(\ref{fip})-(\ref{chip}). The two red diamonds show an example of modes that are in resonance,
i.e. they have the same frequency and same $k_\parallel$. The black line indicates the modes that are resonant with an exactly parallel mode, $k_\perp=0$, 
and divides the space in resonant and non-resonant regions: all the modes lying above the black line do not resonate with a whistler mode. 
The letters indicate the initial dominant mode excited for different runs (see Table \ref{table1}). }\label{fig:resonance}
\end{figure}

\begin{figure}
\noindent\includegraphics[width=20pc]{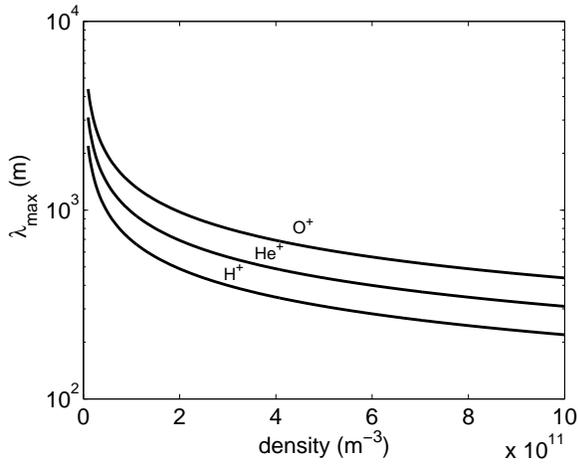}
 \caption{Maximum wavelength $\lambda_{max}$ (in meters) of an exactly parallel whistler wave that can be generated by mode conversion, as a function of density (in $m^{-3}$). 
The three curves are for Oxygen, Helium, and Hydrogen ions.   }\label{fig:lambda}
\end{figure}

We study Eqs. (\ref{fi2})-(\ref{chi2}) in a two-dimensional double-periodic domain $[0,L_x]\times[0,L_y]$.
The best way to understand and characterize the mode coupling due to the inhomogeneity in the density is to transform the equations in Fourier space.\\
Each of the five scalar fields $\phi,V_z,\psi,B_z,\chi$ is decomposed into its Fourier components:
\begin{equation}
 h(x,y,t) = \sum^\infty_{n=-\infty} \sum^\infty_{m=-\infty} \hat{h}_{nm}(t)\exp\left[\frac{2\pi i nx}{L_x}\right]\exp\left[\frac{2\pi i my}{L_y}\right],
\label{DFT}
\end{equation}
where $h$ represents a generic unknown of the system, $m,n$ are integers, and $k_\parallel = \frac{2\pi n}{L_x}, k_\perp = \frac{2\pi m}{L_y} $. 
By inserting Eq.(\ref{DFT}) into Eqs. (\ref{fi2})-(\ref{chi2}), multiplying by $e^{\frac{-2\pi i\alpha x }{L_x}} e^{\frac{-2\pi i\beta y }{L_y}}$, and by using the orthogonality property of the exponential function:
\begin{equation}
 \int_0^L e^{\frac{2\pi i (n-\alpha)x}{L}} dx = L\delta_{n,\alpha} \quad \mbox {   for all  } n,\alpha \in \mathbb{Z},
\end{equation}
one can derive an infinite set of linear ordinary differential equations for the Fourier components $\hat{h}_{nm}(t)$:
\begin{eqnarray}
 -4\pi^2\left(\frac{n^2}{L_x^2} + \frac{m^2}{L_y^2}\right)\frac{d \hat{\varphi}_{nm}}{d t} -\frac{\omega_{ci}}{\omega_{lh}}
\frac{d \hat{B}_{z,nm}}{d t}&=&-\frac{\omega_{ci}}{\omega_{lh}} \frac{2\pi in}{L_x}\hat{V}_{z,nm}, \label{fiDFT} \\
 \frac{d \hat{V}_{z,nm}}{d t}+\frac{\omega_{ci}}{\omega_{lh}} \frac{d \hat{\psi}_{nm}}{d t}&=&
\frac{\omega_{ci}}{\omega_{lh}}2\pi i\left(\frac{n }{L_x}\hat{\varphi}_{nm} -\frac{m}{L_y}\hat{\chi}_{nm}\right), \label{vzDFT} 
\end{eqnarray}
\begin{multline}
 \sum_{\hat{m}=-\infty}^\infty  C_1^{\hat{m}-m} \left[\frac{d \hat{\psi}_{n \hat{m}}}{d t} + 2\pi i \left(\frac{\hat{m}}{L_y}\hat{\chi}_{n \hat{m}} -\frac{n}{L_x}\hat{\varphi}_{n \hat{m}}
\right)\right]  + 4\pi^2\left(\frac{n^2}{L_x^2} + \frac{m^2}{L_y^2}\right)\frac{d\hat{\psi}_{nm}}{dt} =\\
=-\frac{\omega_{ce}}{\omega_{lh}}\frac{2\pi i n}{L_x} \hat{B}_{z,nm}, \label{psiDFT} 
\end{multline}
\begin{multline}
\sum_{\hat{m}=-\infty}^\infty  \left[C_2^{\hat{m}-m}\frac{d \hat{B}_{z,n \hat{m}}}{d t}+
  C_1^{\hat{m}-m}  4\pi^2\left(\frac{n^2}{L_x^2} + \frac{\hat{m}^2}{L_y^2}\right)\frac{d \hat{B}_{z,n \hat{m}}}{d t}+
 C_3^{\hat{m}-m} \frac{2\pi i \hat{m}}{L_y}
\frac{d \hat{B}_{z,n \hat{m}}}{d t }\right]= \\
= \sum_{\hat{m}=-\infty}^\infty \left[\frac{2\pi i n}{L_x}C_2^{\hat{m}-m}
\hat{V}_{z,n \hat{m}}-\frac{\omega_{ce}}{\omega_{lh}} \frac{2\pi i n}{L_x}C_1^{\hat{m}-m}4\pi^2\left(\frac{n^2}{L_x^2} + \frac{\hat{m}^2}{L_y^2}\right)
\hat{\psi}_{n,\hat{m}}\right], \label{bzDFT} 
\end{multline}
\begin{multline}
\sum_{\hat{m}=-\infty}^\infty  \left[-\frac{\omega_{ci}}{\omega_{lh}}  C_3^{\hat{m}-m}  \frac{2\pi i n}{L_x}\frac{d\hat{B}_{z,n \hat{m}}}{dt}
 -C_2^{\hat{m}-m} 4\pi^2\left(\frac{n^2}{L_x^2} + \frac{\hat{m}^2}{L_y^2}\right)\frac{d\hat{\chi}_{n \hat{m}}}{d t}\right]= \\
=\sum_{\hat{m}=-\infty}^\infty  \left[C_1^{\hat{m}-m} 4\pi^2\left(\frac{n^2}{L_x^2} + \frac{\hat{m}^2}{L_y^2}\right)\frac{2\pi i \hat{m}}{L_y}\hat{\psi}_{n \hat{m}}
- C_3^{\hat{m}-m}  4\pi^2\left(\frac{n^2}{L_x^2} + \frac{\hat{m}^2}{L_y^2}\right) \hat{\psi}_{n \hat{m}}\right], \label{chiDFT}
\end{multline}
where the coupling coefficients $C_1^{\Delta m}, C_2^{\Delta m}, C_3^{\Delta m}$ are defined as:
\begin{eqnarray}
 C_1^{\Delta m} = \frac{1}{L_y}\int^{L_y}_0 n_{eq}(y) \exp{\left[\frac{2\pi i \Delta my}{L_y}\right]} dy,\\
 C_2^{\Delta m} = \frac{1}{L_y}\int^{L_y}_0 n_{eq}^2(y) \exp{\left[\frac{2\pi i \Delta my}{L_y}\right]} dy,\\
 C_3^{\Delta m} = \frac{1}{L_y}\int^{L_y}_0 n_{eq}'(y) \exp{\left[\frac{2\pi i \Delta my}{L_y}\right]} dy.
\end{eqnarray}
We note that for a constant $n_{eq}$ these coefficients become $C_1^{\Delta m}=C_2^{\Delta m}=\delta_{\Delta m,0}$ and $C_3^{\Delta m}=0$, and therefore the infinite summations in Eqs. (\ref{fiDFT})-(\ref{chiDFT}) 
reduce to a single term for the $(n,m)$ mode.
This leads to a set of equations that is equivalent to Eqs. (\ref{fip})-(\ref{chip}). In other words, in linear theory mode coupling can only occur in the presence of inhomogeneities.
In addition, since in this paper we consider an equilibrium density which is homogeneous in the $x$ direction, modes with different $n$ wavenumbers (i.e. different $k_\parallel$) remain uncoupled: 
the mode conversion process conserves the parallel wavevector $k_\parallel$. 
By applying a similar procedure using the discrete Fourier transform in the time domain, it is straightforward to see that the 
mode conversion process also conserves the frequency $\omega$ associated with a given $(k_\parallel,k_\perp)$ wavevector. This is due to the fact that the density striation is time-independent.\\
In summary, by applying the discrete Fourier transform to Eqs.(\ref{fi2})-(\ref{chi2}) in space and time separately, one can derive the general rules that govern mode conversion: 
a mode initialized in the homogeneous plasma region with frequency $\omega$ and wavevector $(k_\parallel,k_\perp)$ will be coupled to any mode with the same frequency and the same parallel wavevector.\\
Hence, when a wave-packet encounters a density striation, the dispersion relation for the homogeneous plasma [Eqs. (\ref{fip})-(\ref{chip})] provides the information about which mode will be coupled.\\
Figure \ref{fig:resonance} shows the contour plot of the frequency $\omega$, as a function of $k_\parallel$ and $k_\perp$, derived from the homogeneous plasma dispersion relation [i.e. the solution of Eqs. (\ref{fip}-\ref{chip})] 
for a plasma defined by the ratio of the Alfven velocity to the speed of light $v_A/c= 5\times 10^{-3}$. \\
In this paper we will refer to lower hybrid or to whistler modes, according to the sign of their group velocity in the perpendicular direction. Specifically, we will call lower hybrid the 
modes for which $\frac{\partial \omega}{\partial k_\perp}\cdot k_\perp<0$, and whistler the one for which $\frac{\partial \omega}{\partial k_\perp}\cdot k_\perp>0$. For example, in the parameter regime used for 
Figure \ref{fig:resonance}, the lower hybrid modes correspond roughly to modes with $|k_\perp|>1$. 
The modes that are resonant lie on the intersections of $\omega=const$ and $k_\parallel=const$ curves. An example of such resonant modes is indicated with diamonds in Figure \ref{fig:resonance}. The letters label different numerical simulations (discussed in the following section), which are listed in Table 1.\\
Figure \ref{fig:resonance} shows only the region with $k_\parallel>0$ and $k_\perp>0$. Of course, the dispersion relation is symmetric with respect to the $k_\parallel=0$ and $k_\perp=0$ axis. 
A lower hybrid mode with  $k_\parallel>0$ and $k_\perp>0$ can therefore be resonant with three distinct modes: two whistlers (one with $k_\perp>0$, the other with $k_\perp<0$), and one lower hybrid mode (with $k_\perp<0$). 
Naturally, in Eqs.(\ref{fiDFT})-(\ref{chiDFT}) any mode is also coupled  with itself (self-coupling). 
In the following we will refer to the whistler mode with $k_\perp>0$ as $WH^+$, and to the  whistler mode with $k_\perp<0$ as $WH^-$.The same notation is used for the lower hybrid modes.\\
The black line in Figure \ref{fig:resonance} denotes the modes that are coupled with an exactly parallel propagating whistler mode ($k_\perp=0$). 
It is interesting to notice that modes lying above the black curve are not resonant with a whistler mode: in fact, curves $\omega=const$ and $k_\parallel=const$ 
relative to modes lying above the black curve do not intersect in the region $k_\perp>0$ below the curve. 
Hence, the black line in Figure \ref{fig:resonance} defines the region where mode conversion between lower hybrid and whistler waves is possible.
The black line presents a vertical asymptote at small $k_\parallel$. This means that there is a minimum value of $k_\parallel$ below which there is no mode conversion between  
lower hybrid and an exactly parallel whistler mode. 
The minimum value of $k_\parallel$ corresponds approximately to the value for which the frequency of the parallel propagating whistler is equal to the lower hybrid frequency $\omega_{lh}$:\\
\begin{equation}
\omega_{ce}k_\parallel^2\approx 1
\label{min_k}
\end{equation}
From Eq. (\ref{min_k}), one can derive the maximum wavelength of an exactly parallel whistler wave generated by mode conversion, which is (in dimensional units):\\
\begin{equation}
 \lambda_{max} \approx \frac{3.34\times10^7}{\sqrt{n_{bg}}}\left(\frac{m_i}{m_e}\right)^\frac{1}{4},
\end{equation}
where here $n$ is the density expressed in $m^{-3}$.\\
As expected, $\lambda_{max}$ depends on the value of the density and on the mass of the ions. We show in Figure \ref{fig:lambda} the value of $\lambda_{max}$ as a function of density, for different ions.\\

\section{Numerical results}\label{cp}
We solve numerically the linear set of equations (\ref{fiDFT})-(\ref{chiDFT}), on a grid of size $L_x=3$ km, $L_y=600$ m, and with typically $N_x=N_y=128$ cells in the $x$ and $y$ directions.
The numerical procedure is effectively a spectral method, where the infinite summations of equations (\ref{fiDFT})-(\ref{chiDFT}) are replaced with truncated sums ranging  between $n=-N_x/2,\ldots, N_x/2-1$, and  $m=-N_y/2,\ldots, N_y/2-1$. 
The adimensional form of the equations requires to specify only two parameters: the mass ratio, and the ratio  of the Alfven velocity to the speed of light $v_A/c$.
We focus on hydrogen ions, so that the ion-electron mass ratio is $m_i/m_e=1836$, and we set $v_A/c = 5\times 10^{-3}$. This value is consistent both with observations in the plasmasphere (at an altitude of about 1500 km)
and with the values of density and magnetic field observed in the ionospheric F region (although here the oxygen ions are dominant, and therefore the mass-ratio should be larger). 
Note, however, that the value of the Alfven velocity changes dramatically, by two order of magnitudes, between altitudes of 1000 and 4000 km \citep{lysak99}. For comparison, the same value of $v_A/c$ is used in  \citet{eliasson08}.
The time step is $\Delta t = 0.0025\omega_{lh}^{-1}$, and we have verified that it is small enough so that the total energy is conserved within 0.5\% of the initial energy for most of the simulation runs.
The equations are discretized  in time with a second order implicit backward finite difference (BDF2) scheme \citep{butcher_book}.\\
 For all the runs presented in this paper, the density profile is chosen as 
\begin{equation}
n_{eq}=1-\delta n e^{-\frac{(y-L_y/2)^2}{D_{str}^2}},
\end{equation}
where  $\delta n$ and $D_{str}$ are the amplitude and the width of the striation, respectively.
Consequently, the coupling coefficients become:
\begin{eqnarray}
C_1^{\Delta m}&=&
\begin{cases}
 1 - \delta n \frac{D_{str}}{L_y} \sqrt{\pi}\mathrm{Erf}\left(\frac{L_y}{2D_{str}}\right) &   \quad \mbox{for $\Delta m=0$}\\
\frac{i(\mathrm{e}^{-2\pi i \Delta m} - 1) }{2\pi \Delta m} - \delta n \frac{D_{str}}{L_y}\sqrt{\pi}\mathrm{e}^{-\pi \Delta m\left(i + \frac{D_{str}^2\pi \Delta m}{L_y^2}\right)} & \quad \mbox{for $\Delta m>0$} \label{c1} \\
\end{cases} \\
C_2^{\Delta m}&=&
\begin{cases}
1 + \delta n \frac{D_{str}}{2L_y}\sqrt{\pi}\left[ \sqrt{2}\delta n\mathrm{Erf}\left(\frac{L_y}{\sqrt{2}D_{str}} \right) -4\mathrm{Erf}\left(\frac{L_y}{2D_{str}} \right)\right] & \quad \mbox{for $\Delta m=0$}\\
\delta n \frac{D_{str}}{2L_y}\sqrt{\pi} \left(\sqrt{2}\delta n \mathrm{e}^{-\frac{1}{2}\pi \Delta m \left(2i + \frac{D_{str}^2\pi \Delta m}{L_y^2}\right)} - 4\mathrm{e}^{-\pi \Delta m \left(i+\frac{D_{str}^2\pi \Delta m}{L_y^2}  \right)} \right) & \quad \mbox{for $\Delta m>0$} \label{c2} \\
\end{cases}\\
C_3^{\Delta m}&=&
\begin{cases}
0 &   \quad \mbox{for $\Delta m=0$}\\
-2i\Delta m\delta n\frac{D_{str}}{L_y^2}\pi^{\frac{3}{2}}\mathrm{e}^{-\pi \Delta m \left(\frac{D_{str}^2\pi \Delta m}{L_y^2} -i  \right)}& \quad \mbox{for $\Delta m>0$} \label{c3}
\end{cases}
\end{eqnarray}

\begin{figure}
 \includegraphics[width=35pc]{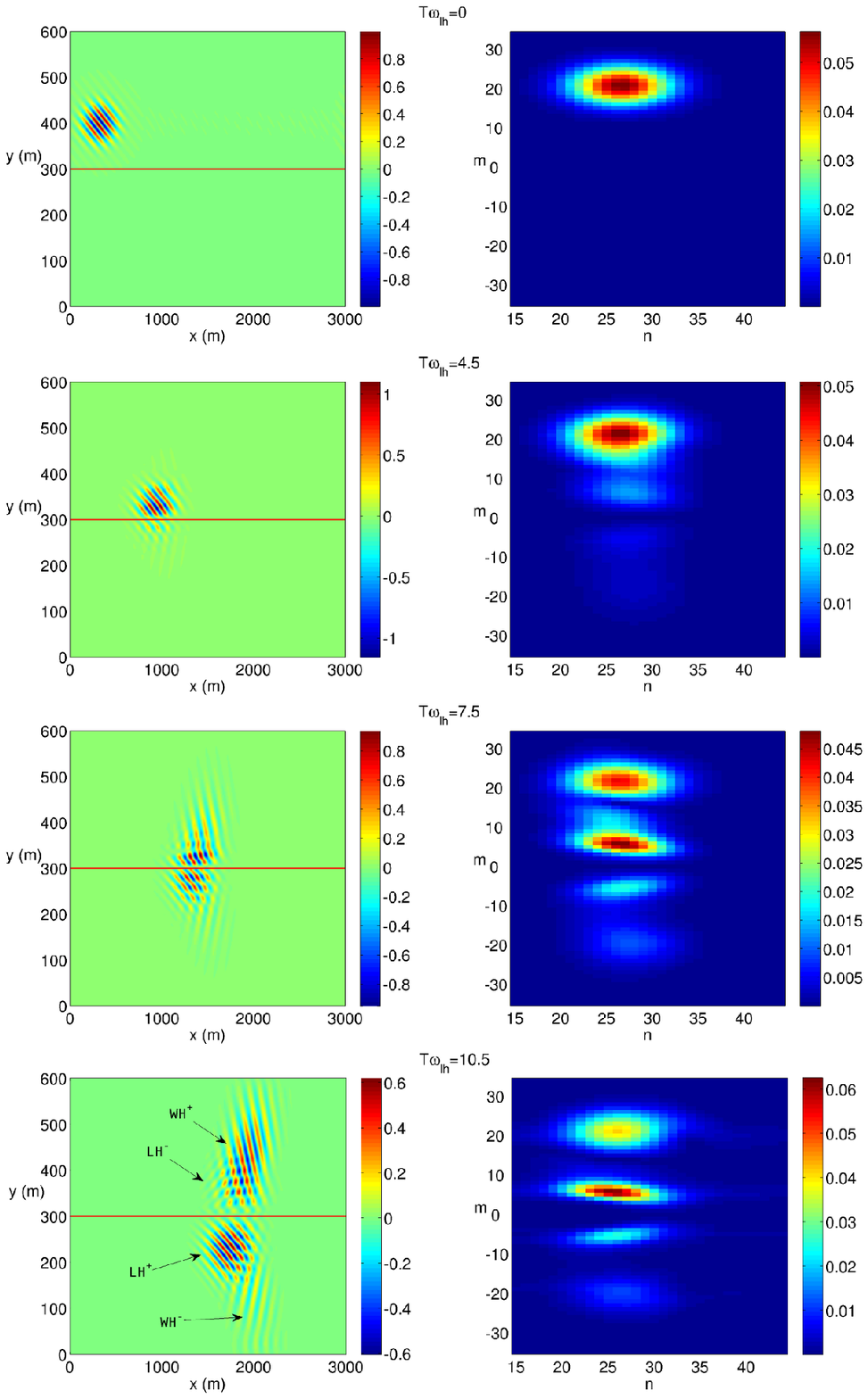}
\caption{Evolution of the initial dominant mode $(n_0=26, m_0=21)$, for $\delta n=0.5$ and $D_{str}=1.2$ at four times $T\omega_{lh}=0, 4.5,7.5, 10.5$. 
Left panels: $B_x$ in physical space. Right panels: amplitude of the Fourier modes of $B_x$. 
For time $T\omega_{lh}=10.5$  four arrows indicate the whistler and lower hybrid wave packets generated through the mode conversion process.}\label{fig:26_21_1}
 \end{figure}

\begin{figure}
\noindent\includegraphics[width=20pc]{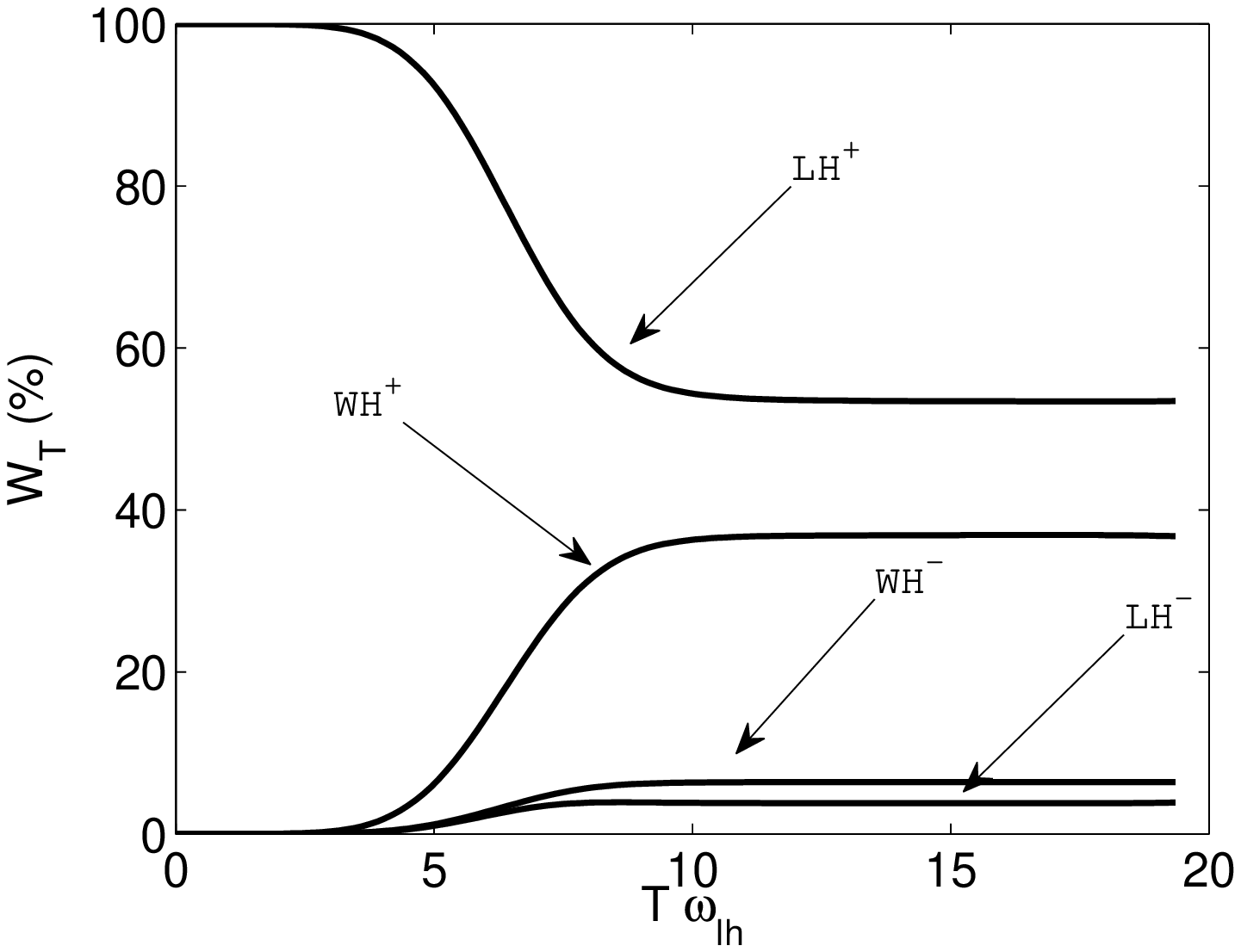}
\caption{Total energy $W_T=W_E+W_B+W_K$ in time, for the case with initial dominant modes $(n_0=26, m_0=21)$, $\delta n=0.5$, and $D_{str}=1.2$. The four curves indicate the partition of the total energy among the four wave packets.}\label{fig:energy1}
\end{figure}

\begin{figure}
\noindent\includegraphics[width=20pc]{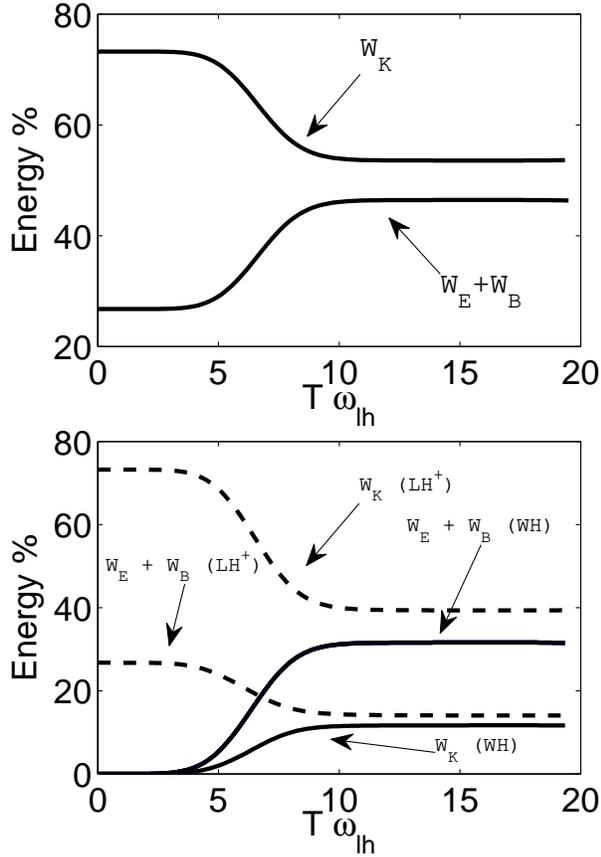}
\caption{Initial dominant mode $(n_0=26, m_0=21)$, $\delta n=0.5$, and $D_{str}=1.2$. 
Top panel:  kinetic $(W_K)$ and electromagnetic $(W_E+W_B)$ energies in time, computed over the whole domain (i.e. summed over all the modes). 
Bottom panel: partition of the kinetic and electromagnetic energies between $LH^+$ and $WH=WH^++WH^-$.}\label{fig:energy2}
\end{figure}

At time $T =0$, a lower hybrid packet is initialized in wavevector space exciting $25^2$ modes in $(k_\parallel,k_\perp)$, centered around a dominant mode $(n_0,m_0)$.
The amplitude of each mode of the packet is chosen such that the amplitude of the scalar field $\chi$ is given by the Gaussian:
\begin{equation}\label{gaussian}
\exp(-(n-n_0)^2/\alpha - (m-m_0)^2/\beta),
\end{equation}
with $\alpha=0.0072$, and $\beta=0.1436$, and the amplitudes of the other scalar fields follow from the linear dispersion relation [Eqs. (\ref{fip})-(\ref{chip})]. \\
The total energy of the system in physical space is defined as 
\begin{equation}
 W= \frac{1}{2}\int_V \left(\frac{\mathbf{B}\cdot\mathbf{B}}{\mu_0} + \varepsilon_0\mathbf{E}\cdot\mathbf{E} \right)dV +\sum_{s=i,e } \int_V \frac{m_s}{2}n_{eq}\mathbf{V}_s^2 dV
\end{equation}
where the integrals are over the whole domain, $\varepsilon_0$ is the vacuum permittivity, and the higher order contributions in the kinetic energy have been neglected.
The energy associated with a mode $(n,m)$ in Fourier space are defined, after normalization, as:
\begin{eqnarray}
W_B^{n,m} &=& |\hat{B}_{x,nm}|^2 + |\hat{B}_{y,nm}|^2 + |\hat{B}_{z,nm}|^2 \\
W_E^{n,m} &=& \left(\frac{\omega_{lh}}{\omega_{pe}}\right)^2  \left(|\hat{E}_{x,nm}|^2 + |\hat{E}_{y,nm}|^2 + |\hat{E}_{z,nm}|^2\right) \\
W_K^{n,m} &=& \sum_{s=i,e }\frac{n_0m_s}{B_0^2\varepsilon_0}\left(\frac{\omega_{lh}}{\omega_{pe}}\right)^2 \left( |\hat{w}_{x,nm}^{s}|^2 + |\hat{w}_{y,nm}^{s}|^2+ |\hat{w}_{y,nm}^{s}|^2\right),
\end{eqnarray}
where $W_B, W_E$, and $W_K$ denote magnetic, electric, and total kinetic energy, respectively. The relationship between vector fields and the scalar fields are reported in the Appendix.
We define the efficiency $\eta$ of the mode conversion process as the ratio of the energy residing in the whistler modes (with positive and negative $k_\perp$), over the energy in $LH^+$ at $T=0$:
\begin{equation}
\eta=\frac{\sum\limits_ {n,m=WH^+, WH^-} \left(W_B^{n,m}+W_E^{n,m}+W_K^{n,m}\right)}{\sum\limits_ {n,m=LH^+} \left(W_B^{n,m}+W_E^{n,m}+W_K^{n,m}\right)},\label{eff_formula}
\end{equation}
where the summation at numerator is over the modes belonging to either the $WH^+$ and $WH^-$ wave packet, and the summation at denominator is over modes belonging to $LH^+$ (for the runs shown in Section 3.1 and 3.2 , the total initial energy is 
contained in $LH^+$).\\
The efficiency of the mode conversion depends on three factors: the values of $ \delta n$ and $D_{str}$, and the initial condition. 
Finding the maximum conversion efficiency would require the study of \emph{all} possible initial states and it is not the goal of this paper. 
Therefore, in Sections (\ref{dstr}) and (\ref{deltan})  we approach this problem not focusing on the 
absolute value of the efficiency, but rather on its scaling properties with respect to changes in $ \delta n$ and $D_{str}$. 
In Section (\ref{IC}) we will perform some studies changing the initial condition in an attempt to improve the efficiency of mode conversion.\\
Firstly, we show the typical time evolution of the mode conversion process in Figure \ref{fig:26_21_1}. 
In this case the initial mode is $(n_0=26, m_0=21)$ (corresponding to $k_\parallel=0.41$ and $k_\perp=1.654$), $\delta n=0.5$, and $D_{str} = 1.2$.  
Note that this mode is not purely electrostatic.
Figure \ref{fig:26_21_1}  shows four snapshots in time of the parallel component of the magnetic field, $B_x$, in physical space (left panels), 
and the amplitude of its Fourier transform in $k$-space (right panels), in arbitrary units. The horizontal red line in the left panels  shows the location of the minimum of the equilibrium density. 
At time $T=0$ the wave packet is initialized at an approximate distance of 13 $d_e$ from the striation, with the procedure described above. 
The packet moves downwards at its group velocity, and encounters the striation at around $T\omega_{lh}=4.5$. 
The mode conversion proceeds as the initial $LH^+$ packet passes through the striation. 
At time $T\omega_{lh}=10.5$ it can be seen that two whistler modes ($WH^+$ and $WH^-$) and one lower hybrid mode ($LH^-$) have been generated.
 Each wave packet moves with its own group velocity: $LH^+$ and $WH^-$ move downwards, while $LH^-$ and $WH^+$ move upwards. 
The whistlers are generally faster than the lower hybrid modes, their angle of propagation is more parallel, and their wave front is quasi-perpendicular to the background magnetic field, 
so that one can easily identify the four different modes in physical space, which are denoted with arrows in  Figure \ref{fig:26_21_1}.\\
Figure \ref{fig:energy1} shows the total energy evolution in time for the four different wave packets, normalized to the total energy of the system. 
The total energy is very well conserved with a maximum loss of only 0.2\% during the simulation. At the end of the run $LH^+$ (which initially had 100\% of the energy) has lost 47\% of its energy, 
and $WH^+$, $WH^-$, and $LH^-$ account respectively for 37\%, 6\%, and  4\% of the total energy. Hence, in this case the efficiency is 43\%, 
and almost 80\% of the energy lost by the initial wave packet has been converted to energy in $WH^+$. 
We will see that there are other cases where, by changing the value of $D_{str}$, the final partition of energy is spread more equally between $WH^+$ and $WH^-$.\\
Figure \ref{fig:energy2} shows the contribution of kinetic ($W_K$) and electromagnetic ($W_E+W_B$) energy to the total energy, for the same run of Figs. \ref{fig:26_21_1} and \ref{fig:energy1}. 
The top panel represents the energies calculated in the whole box (i.e. summed over all the modes). 
Note that the mode conversion process is characterized by a transfer of energy from kinetic to electromagnetic: the former changes from 73\% to 58\%, and the latter changes from 27\% to 42\%.  
This happens because the kinetic energy is larger than the electromagnetic energy for $LH^+$, while it is smaller for whistler modes. This can be seen in the bottom panel of Figure \ref{fig:energy2}. 
The energy of the initial lower hybrid mode is composed of about 73\% kinetic and 27\% electromagnetic energies, 
while at the end of the simulation the partition of energy for the whistler modes (here computed as the sum of $WH^+$ and $WH^-$) is about 28\% kinetic and 72\% electromagnetic 
(they account respectively for 12\% and 31\% of the total energy).\\
From Figs. \ref{fig:energy1} and \ref{fig:energy2} one can see that the mode conversion process has a duration of about $T\omega_{lh}=5$, which is approximately the transit time of the initial lower hybrid packet through the striation.
Once $LH^+$ has passed the striation, the modes stop converting, and the energies remain flat for the remainder of the simulation, provided that the simulation box is large enough so that waves do not escape and re-enter from the boundaries.\\
We have run several simulations, changing the position in $k$-space of the dominant mode $(n_0,m_0$), and the value of $\delta n$ and $D_{str}$.
Table (\ref{table1}) presents the maximum conversion efficiency achieved for various initial dominant modes, along with the error in energy conservation (defined as the maximum variation of the total energy during the simulation:
$\Delta W_T = \mathrm{max} |W_T-W_T(T=0)|/|W_T(T=0)|$), 
for cases with $\delta n=0.5$, and $D_{str}=1$. The last column indicates the angle of propagation of the generated whistler mode. The initial modes are plotted in Figure \ref{fig:resonance} and labeled with letters.
A general trend is that the efficiency is higher when $m_0$ is lower. This can be explained by the fact that the absolute value
of the coupling coefficients $C_1^{\Delta m}$ and $C_2^{\Delta m}$ in Eqs. (\ref{c1}),(\ref{c2}) are decreasing monotonically as $\Delta m$ increases, which means that the stronger coupling happens when the perpendicular wavenumbers 
corresponding to the modes $LH^+$ and $WH^+$ are closer. If one looks at the runs performed along $\omega=const$ curves (i.e. the set of runs A-B-C-E-G, D-F-H-I-J, and L-N-O-P), one can note that the efficiency is much higher when $m_0$ 
is smaller. For instance, $\eta$ goes from 43\% for case A ($m_0=20$) to 1.4\% for case G ($m_0=48$). Similarly, $\eta$ changes from 40\% for case D ($m_0=17$) to 7\% for case J ($m_0=38$), and from 49\% for case L ($m_0=17$)
to 18.5\% for case P ($m_0=27$). 
Also note that when the initial mode is resonant with an exactly parallel whistler (the black line of Figure \ref{fig:resonance} and the case treated by \citet{eliasson08} for $k_\perp\gg k_\parallel$), 
the efficiency is affected by the fact that some modes that are in the non-resonant part of the $(k_\parallel,k_\perp)$ space 
are also initially excited (see Figure \ref{fig:resonance}). 
While our maximum efficiency for all these runs is 50\% (and 25\% for conversion to an exactly parallel propagating whistler), we will see in the next section that a careful choice of the value of $D_{str}$ can have a large effect on the efficiency achieved.
\subsection{Dependence of the conversion efficiency on the striation width}\label{dstr}
\begin{figure}
\noindent\includegraphics[width=20pc]{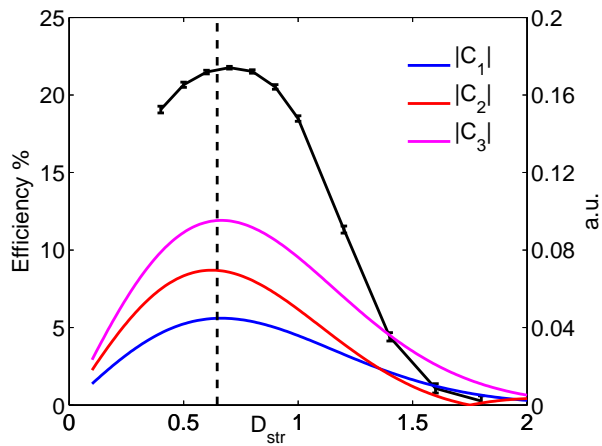}
 \caption{Initial dominant mode: $(n_0=30,m_0=27)$. The solid black line is the maximum efficiency (vertical axis on the left), as a function of $D_{str}$. 
Blue, red and purple lines are respectively the absolute value of the coupling coefficients $C_1^{\Delta m},C_2^{\Delta m},C_3^{\Delta m}$, for $\Delta m=27$, as functions of $D_{str}$, in arbitrary units (vertical axis on the right).
The dashed vertical line denotes the average value of $D_{str}$ for which the three coupling coefficients are maximum. 
For this case the initial mode is resonant with an exactly parallel whistler mode, therefore $WH^+=WH^-$. }\label{fig:eff_30_27}
\end{figure}

\begin{figure}
\noindent\includegraphics[width=20pc]{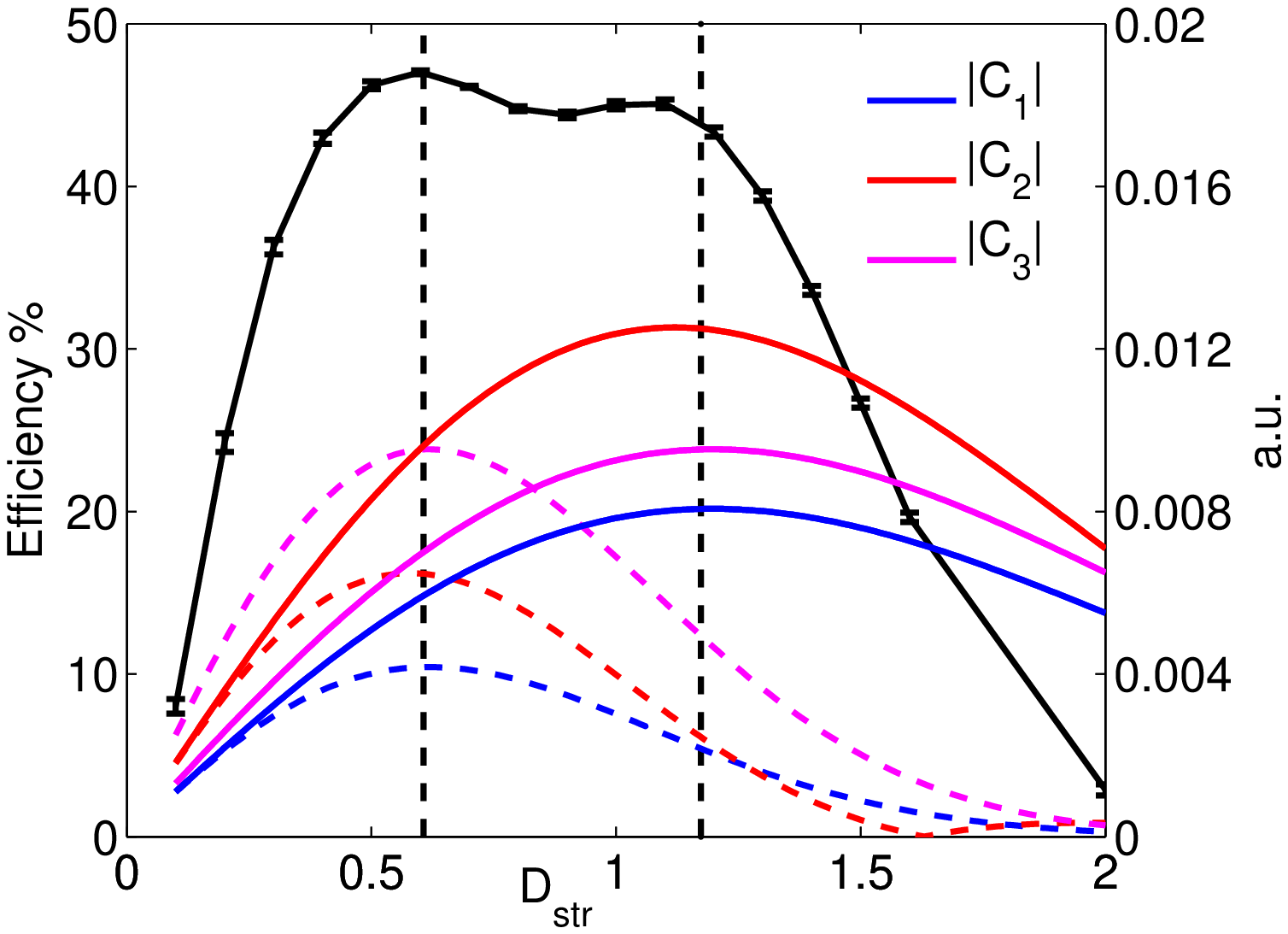}
 \caption{Initial dominant mode: $(n_0=26,m_0=21)$. The solid black line is the maximum efficiency (vertical axis on the left), as a function of $D_{str}$. 
Blue, red and purple lines are respectively the absolute value of the coupling coefficients $C_1^{\Delta m},C_2^{\Delta m},C_3^{\Delta m}$, for coupling with 
$WH^+$ (solid lines, $\Delta m=16$) and $WH^-$ (dashed lines, $\Delta m=27$), as functions of $D_{str}$, in arbitrary units (vertical axis on the right).
The dashed vertical lines denote the average values of $D_{str}$ for which the three coupling coefficients are maximum.}\label{fig:eff_26_21}
\end{figure}

\begin{figure}
 \includegraphics[width=35pc]{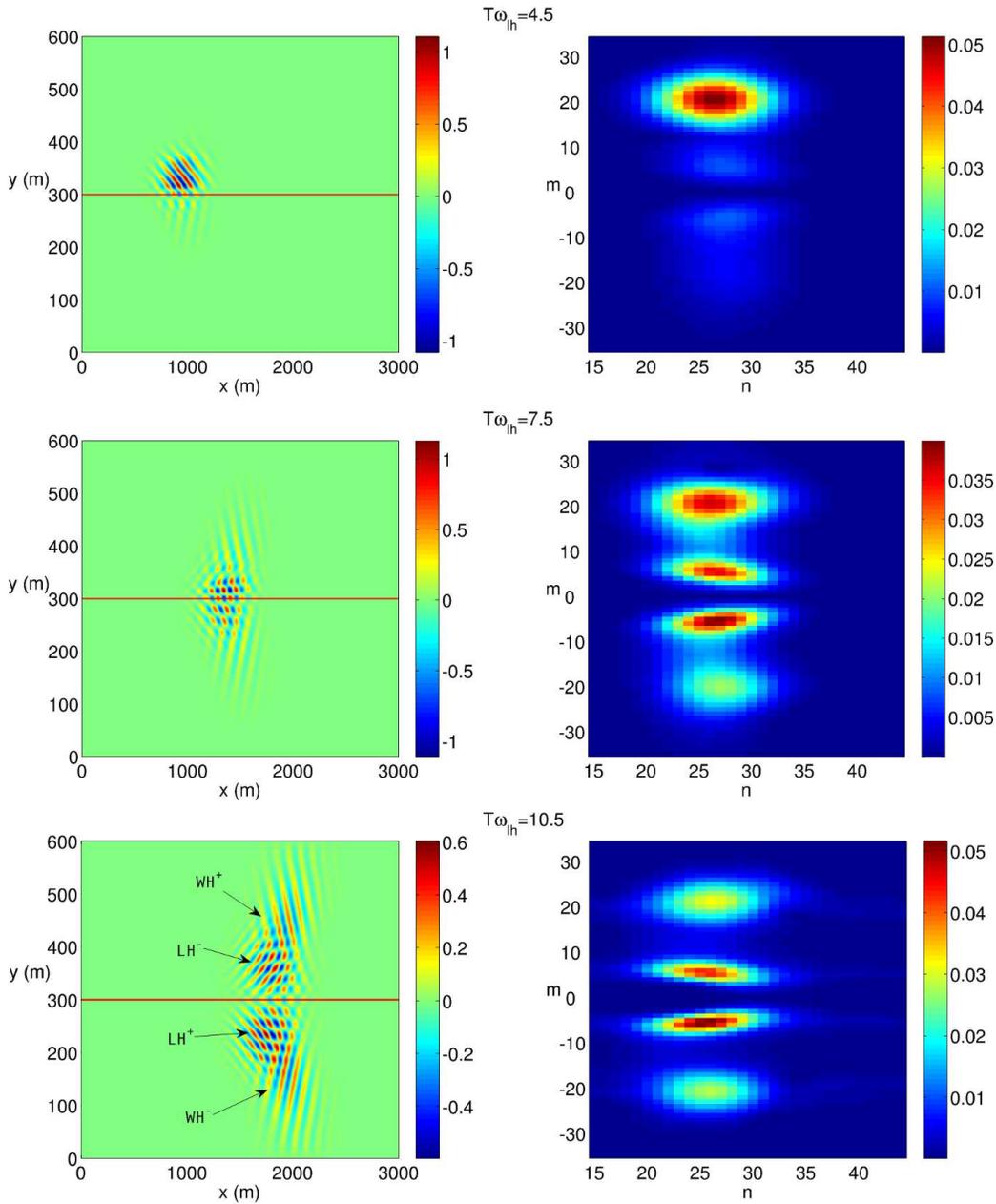}
\caption{Evolution of the initial dominant mode $(n_0=26, m_0=21)$, for $\delta n=0.5$ and $D_{str}=0.6$ at four times $T\omega_{lh}=0, 4.5,7.5, 10.5$. Left panels: $B_x$ in physical space. Right panels: amplitude of the Fourier modes of $B_x$. For time $T\omega_{lh}=10.5$  four arrows indicate the whistler and lower hybrid wave packets generated through the mode conversion process.}\label{fig:26_21_2}
 \end{figure}

\begin{figure}
\noindent\includegraphics[width=20pc]{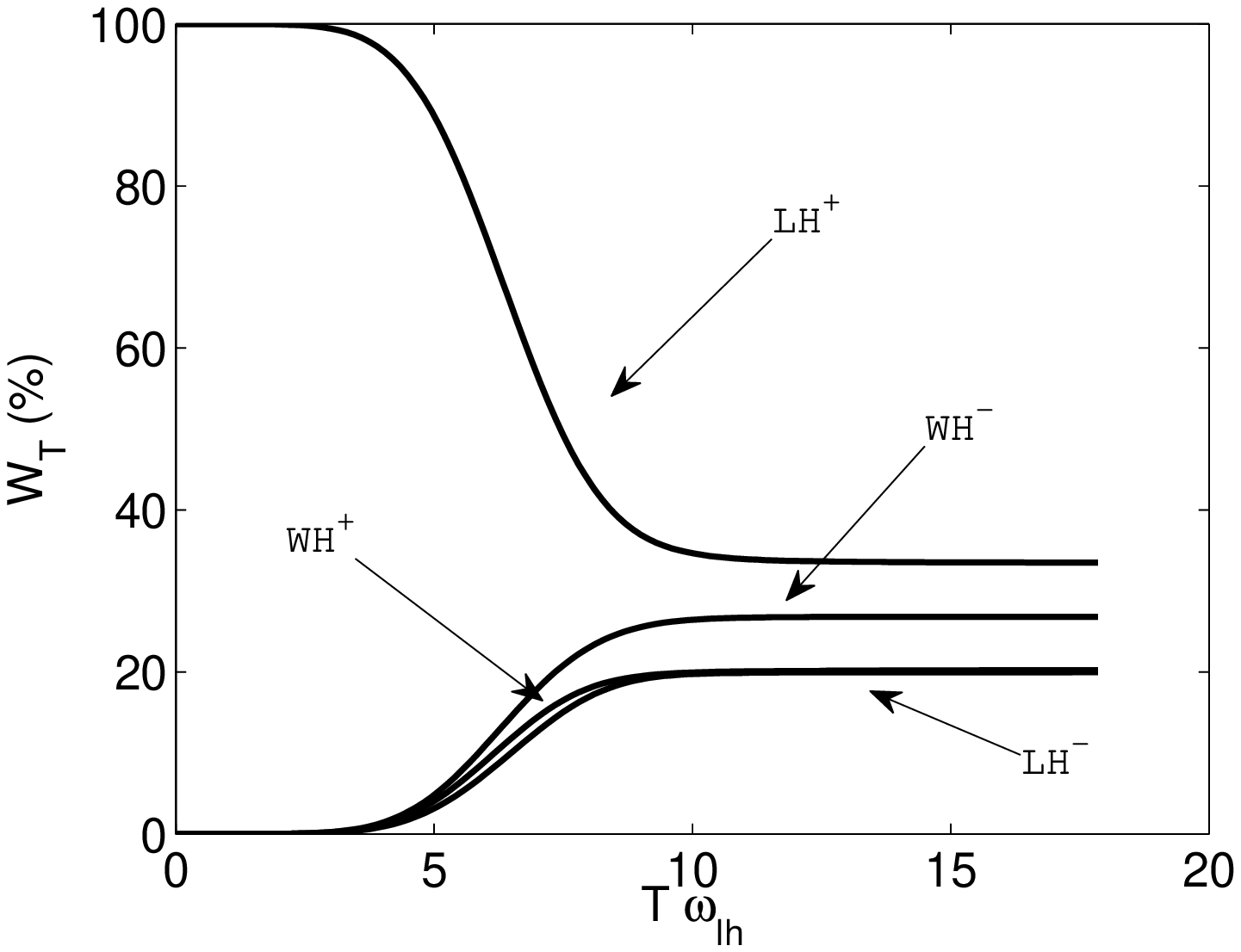}
\caption{Total energy $W_T+W_E+W_B+W_K$ in time, for the case with initial dominant modes $(n_0=26, m_0=21)$, $\delta n=0.5$, and $D_{str}=0.6$. The four curves indicate the partition of the energy among the four wave packets. }\label{fig:energy3}
\end{figure}

\begin{figure}
\noindent\includegraphics[width=20pc]{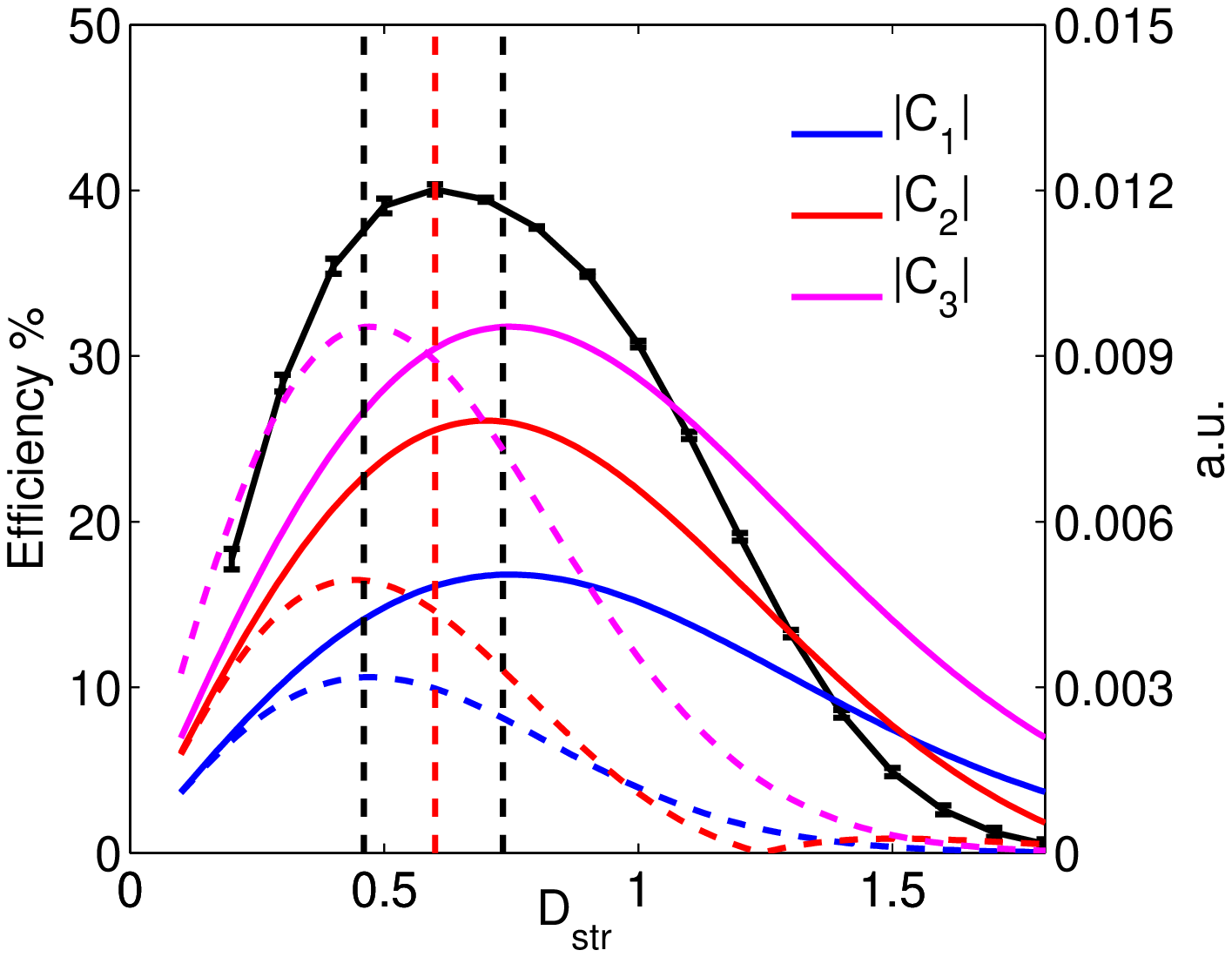}
 \caption{Initial dominant mode: $(n_0=19,m_0=28)$. The black line is the maximum efficiency (vertical axes on the left), as a function of $D_{str}$. 
Blue, red and purple lines are respectively the absolute value of the coupling coefficients $C_1^{\Delta m},C_2^{\Delta m},C_3^{\Delta m}$, for coupling with 
$WH^+$ (solid lines, $\Delta m=24$) and $WH^-$ (dashed lines, $\Delta m=33$), as functions of $D_{str}$, in arbitrary units (vertical axes on the right).
The black dashed vertical lines denote the values of $D_{str}$ that is the average value for which the three coupling coefficients are maximum.
The red dashed vertical line denote the values of $D_{str}$ that is the average value for which the sum of the $WH^+$  and $WH^-$ coupling coefficients are maximum.}\label{fig:eff_19_28}
\end{figure}

\begin{figure}
 \includegraphics[width=35pc]{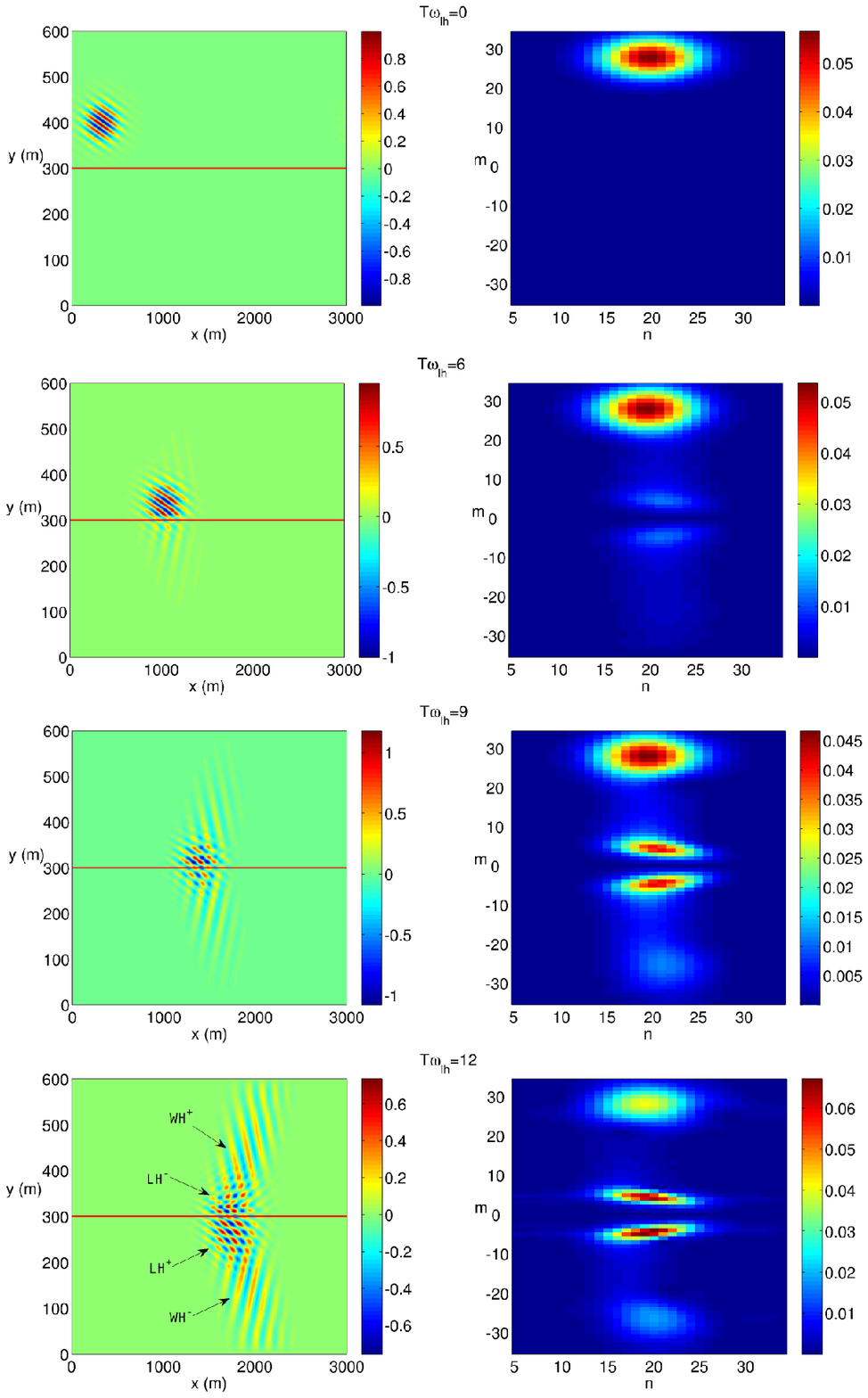}
\caption{Evolution of the initial dominant mode $(n_0=19, m_0=28)$, for $\delta n=0.5$ and $D_{str}=0.6$ at four times $T\omega_{lh}=0, 4.5,7.5, 10.5$. Left panels: $B_x$ in physical space. Right panels: amplitude of the Fourier modes of $B_x$. For time $T\omega_{lh}=10.5$  four arrows indicate the whistler and lower hybrid wave packets generated through the mode conversion process.}\label{fig:19_28}
 \end{figure}

\begin{figure}
\noindent\includegraphics[width=20pc]{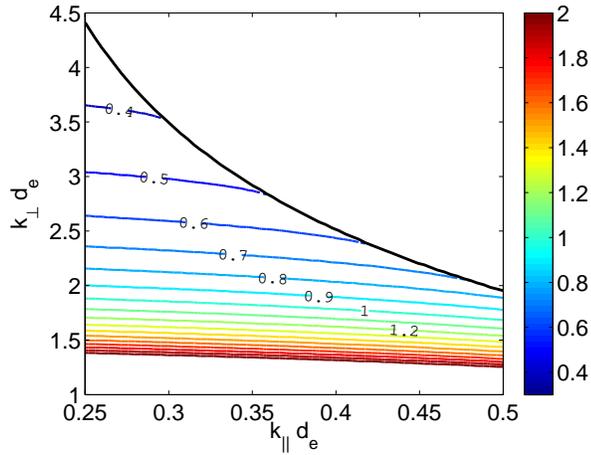}
 \caption{Contour plot of the value of $D_{str}$ that maximizes the conversion efficiency between $LH^+$ and $WH^+$, as a function of the initial mode, in the $(k_\parallel,k_\perp)$ space. The value of $D_{str}$ is calculated by using the criterion in Eq. (\ref{ktimesD}).}\label{fig:Dstr}
\end{figure}

In this section, we investigate how the value of $D_{str}$ affects the conversion efficiency. Figure \ref{fig:eff_30_27} shows the efficiency for the initial mode $n_0=30, m_0=27$  $(k_\parallel=0.473, k_\perp=2.127)$ 
as a function of $D_{str}$, with $\delta n=0.5$. 
The solid black line denotes the maximum efficiency reached during the simulation (vertical scale on the left). The error bar is given by the error in energy conservation. 
Blue, red, and purple lines represent the values of the coupling coefficients $C_1^{\Delta m}, C_2^{\Delta m},C_3^{\Delta m}$ [Eqs.(\ref{c1})-(\ref{c3})], respectively, 
where $\Delta m$ is given by the difference between $m_0$ and the perpendicular wavenumber of the resonant whistler mode. 
Since the mode $(n_0=30, m_0=27)$ is resonant with the exactly parallel mode $(n=30, m=0)$, $\Delta m$ is equal to 27.
 Not surprisingly the highest efficiency ($\eta\approx 22\%$) is achieved when the three coupling coefficients are close to their peak value.
The vertical dashed line indicates the average value of $D_{str}$ for which $C_1^{\Delta m}, C_2^{\Delta m}$,and $C_3^{\Delta m}$ reach their maximum. \\
Figure \ref{fig:eff_26_21} shows the efficiency for the mode $(n_0=26, m_0=21)$, with the same format as in Figure \ref{fig:eff_30_27}. This mode is now resonant with two different whistlers, with wavenumbers $m$ equal to $\pm 5$.
The solid and dashed lines show the value of the coupling coefficients $C_1^{\Delta m}, C_2^{\Delta m},C_3^{\Delta m}$ for $\Delta m=16$ (coupling with $WH^+$), and 
for $\Delta m =27$ (coupling with $WH^-$), respectively.
The average values of $D_{str}$ where the two sets of curves (solid for $WH^+$ and dashed for $WH^-$) reach their maximum are $D_{str}=1.2$, and $D_{str}=0.6$, and they are again indicated by dashed vertical lines. 
Interestingly, the efficiency has now two local maxima, each corresponding to the maximum of each set of the coupling coefficients. 
The $WH^-$ mode will be dominant for $D_{str}=0.6$, while the $WH^+$ mode will be dominant for $D_{str}=1.2$. 
The latter case has already been discussed in Figs. \ref{fig:26_21_1}, \ref{fig:energy1}, and \ref{fig:energy2}, where almost 80\% of the energy converted from $LH^+$ is channeled to $WH^+$.
The case for $D_{str}=0.6$ is shown in Figure \ref{fig:26_21_2}, with the same snapshots in time as in Figure \ref{fig:26_21_1} (except for $T\omega_{lh}=0$). 
The corresponding energy plot is represented in Figure \ref{fig:energy3}. At time $T\omega_{lh}=18$ $LH^+$ has lost about 67\% of its initial energy, 
which is partitioned as the following: 27\% in $WH^-$, 20\% in $WH^+$, and 20\% in $LH^-$. Therefore, in this case the efficiency is $\eta=47\%$, and $WH^-$ dominates over $WH^+$.\\
Another interesting case happens when the coupling coefficients relative to the two 
oppositely propagating whistlers have a maximum for similar values of $D_{str}$. In this case the two peaks of the efficiency merge into a single maximum.
This case is shown in Figure \ref{fig:eff_19_28} where the efficiency for the initial mode $(n_0=19, m_0=28)$ is presented with the same format as in Figure \ref{fig:eff_26_21}. In this case the maxima of the  coupling coefficients relative to $WH^+$ and $WH^-$ are relatively close to each other. 
Once again, the average value of $D_{str}$ where they reach their maximum is indicated with two black dashed vertical lines. The red vertical line indicates the average of the two.
Figure \ref{fig:19_28} shows the time evolution of this case [initial mode $(n_0=19, m_0=28)$], for $D_{str}=0.6$ (i.e. at the maximum efficiency, $\eta\approx40\%$). 
The error in energy conservation is $\Delta W_T=0.3$\%, and the energies at the end of the simulation (not shown) are: 47\% in $LH^+$, 13\% in $LH^-$, 19\% in $WH^+$, and 21\% in $WH^-$.
Hence, the two whistlers have now comparable amplitudes.\\
From the results shown in Figs. \ref{fig:eff_30_27}-\ref{fig:eff_19_28} it is clear that the maximum of the coupling coefficients (\ref{c1})-(\ref{c3}) provides a good estimate of 
the optimal striation width  which will result in the maximum efficiency.
In fact, it is not surprising that when the coupling is stronger, the conversion between modes is more efficient.
In order to derive a general formula for the the value of $D_{str}$ for which the efficiency is highest, one has to calculate the maximum of $C_1^{\Delta m}, C_2^{\Delta m},C_3^{\Delta m}$ with respect to $D_{str}$. 
For $C_1^{\Delta m}$ and $C_3^{\Delta m}$,
this can be done analytically, obtaining the result:
\begin{equation}
 \Delta k_\perp D_{str}=\sqrt{2},\label{ktimesD}
\end{equation}
where $\Delta k_\perp=2\pi\Delta m/L_y$.
For the coefficient $C_2^{\Delta m}$, formula (\ref{ktimesD}) holds only in the limit $\delta n=0$. More generally, the maximum of $C_2^{\Delta m}$ is constrained in the quite narrow range $1.2\lesssim \Delta k_\perp D_{str}\leq \sqrt{2}$, 
and for all practical purposes one can consider the unique criterion 
\begin{equation}\label{criterion}
\Delta k_\perp D_{str}\simeq 1.4.
\end{equation}
We notice that \citet{eliasson08} have empirically found the formula $k_\perp D_{str}\approx 1.5$ as a criterion for the maximum conversion efficiency. 
For the case they have studied, $k_\perp\simeq \Delta k_\perp$, and therefore the two criteria are equivalent.
However, their formula is only valid for the cases in which an exactly parallel whistler is generated. 
On the other hand, our criterion $\Delta k_\perp D_{str}\simeq 1.4$ is generally applicable, and has been derived more rigorously from the mathematical expressions of the coupling coefficients.\\
From the criterion (\ref{criterion}) for maximum efficiency, one can derive the optimal striation width corresponding to a given initial mode.
In Figure \ref{fig:Dstr}, we show the contour plot of the optimal $D_{str}$ in the $(k_\parallel,k_\perp)$ space that will result in the most efficient coupling between $LH^+$ and $WH^+$. 
Note that since we have defined the conversion efficiency as the ratio of the final energy contained in both $WH^+$ and $WH^-$ over the total energy in the box, the value of $D_{str}$ 
that maximizes the conversion between the lower hybrid mode and one whistler might not be equal to the value of $D_{str}$ that results in the maximum efficiency 
(this happens for instance for the case discussed in Figure \ref{fig:eff_19_28}).\\
From Eq. (\ref{criterion}) one can also see that when $k_\perp D_{str}\simeq 0.7$ the conversion between $LH^+$ and $LH^-$ will be most efficient (because in this case $\Delta k_\perp=2k_\perp$).
\subsection{Dependence of the conversion efficiency on the striation amplitude}\label{deltan}
\begin{figure}
 \includegraphics[width=20pc]{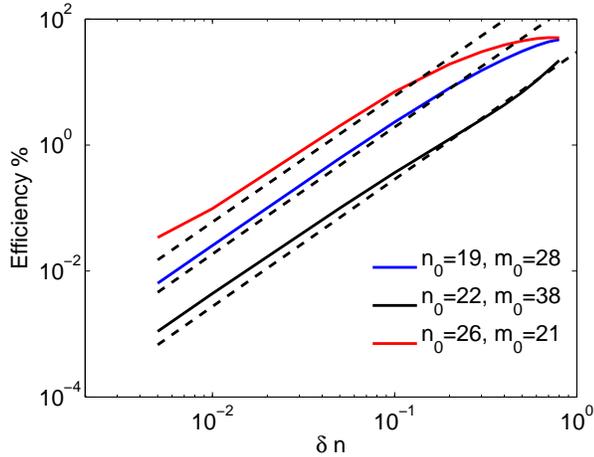}
\caption{Conversion efficiency for the initial dominant modes $(n_0=19, m_0=28), (n_0=22, m_0=38)$, and $
(n_0=26, m_0=21)$, for $D_{str}=1$, as a function of $\delta n$. The power law $y\sim\delta n^a$, with $a=2.016, 2.02, 2.1$ for the modes $(n_0=19,m_0=28), (n_0=22,m_0=38),(n_0=26,m_0=21)$ is indicated with dashed line.}\label{fig:delta_n}
 \end{figure}

In this section, we investigate the role of the striation amplitude on the conversion efficiency. We vary $\delta n$ parametrically up to a density depletion of 80\%. Strong density depletions are relevant, for instance, 
to the recent experiments at LAPD \citep{vanCompernolle11}.\\
The coupling coefficients $C_1^{\Delta m}$ and $C_3^{\Delta m}$ in Eqs. (\ref{c1})-(\ref{c3}) are linear functions of $\delta n$, while the coupling coefficient $C_2^{\Delta m}$ in Eq. (\ref{c2}) 
is a quadratic function of $\delta n$. The three coefficients enter in the model equations (\ref{fiDFT})-(\ref{chiDFT})
in a non-trivial way. As it can be seen from Figs. \ref{fig:eff_30_27}, \ref{fig:eff_26_21}, and \ref{fig:eff_19_28} which one of the three terms dominates is case dependent.
For this reason it is not straightforward to derive a scaling law for the efficiency with respect to $\delta n$, and we have studied it numerically.
Figure \ref{fig:delta_n} shows the change in efficiency, for the cases $(n_0=19,m_0=28), (n_0=22,m_0=38),(n_0=26,m_0=21)$ for $D_{str}=1$, as a function of $\delta n$. 
The error in energy conservation is always much smaller than the efficiency and is not plotted, as it would not be noticeable.
The efficiency is generally higher for larger $\delta n$. For $\delta n \leq 0.1$ all of the three curves can be fitted reasonably well by a power-law:
\begin{equation}
\eta \approx \delta n^a.
\end{equation}
The exponent $a$ has been calculated by a standard Levenberg-Marquardt (nonlinear least-square) method \citep{press}, 
yielding the result $a=2.016, 2.02, 2.1$ for the modes $(n_0=19,m_0=28), (n_0=22,m_0=38),(n_0=26,m_0=21)$, respectively.
The power-law fits are indicated with dashed lines in Figure \ref{fig:delta_n}.
\subsection{The role of the initial condition on the conversion efficiency}\label{IC}
\begin{figure}
 \includegraphics[width=20pc]{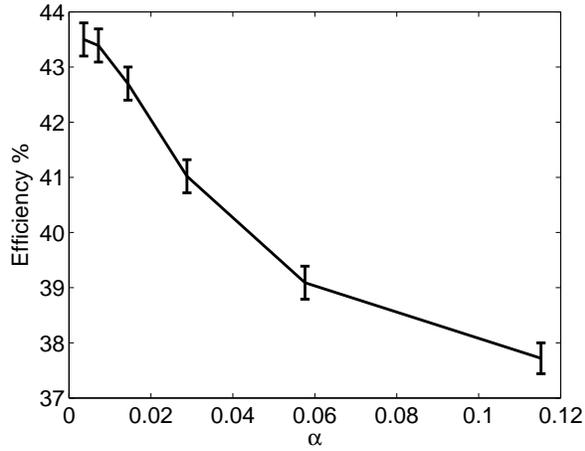}
\caption{Conversion efficiency for the initial dominant mode $(n_0=26, m_0=21)$, $\delta n=0.5$, and $D_{str}=1.2$, as a function of $\alpha$, for $\beta=0.1436$. The error bar is given by the error in energy conservation.}\label{fig:eff_alpha}
 \end{figure}

\begin{figure}
 \includegraphics[width=20pc]{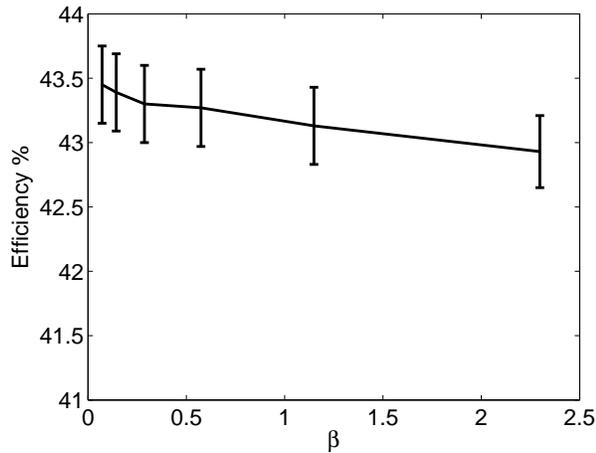}
\caption{Conversion efficiency for the initial dominant mode $(n_0=26, m_0=21)$, $\delta n=0.5$, and $D_{str}=1.2$, as a function of $\beta$, for $\alpha=0.0072$. The error bar is given by the error in energy conservation.}\label{fig:eff_beta}
 \end{figure}

The goal of this section is to show a strategy that allows to increase the conversion efficiency shown in the previous section, and to study the robustness
of the value of the efficiency against changes in $\alpha$ and $\beta$ (i.e. the parallel and perpendicular width of the Gaussian in Eq. (\ref{gaussian})).
As already stated, such efficiency is not only a function of the values of $D_{str}$ and $\delta n$, but also of the particular initial condition chosen. 
In order to address the robustness of the efficiencies reported so far, we show in Figures \ref{fig:eff_alpha} and \ref{fig:eff_beta}, the efficiency for the case with initial dominant mode $(n_0=26, m_0=21)$, $D_{str}=1.2$, $\delta n=0.5$, 
changing $\alpha$ and $\beta$ respectively (keeping the other constant). It can be seen that the efficiency is lower for wider packets in Fourier space, although it is more dependent on $\alpha$ than on $\beta$. 
Indeed, the efficiency changes from 43.5\% to 37.7\% when $\alpha$ varies from 0.0036 to 0.115, while it changes only from 43.5\% to 42.9\% when $\beta$ varies from 0.072 to 2.298. 
The fact that wider packets in Fourier space result in lower conversion efficiencies can be interpreted by the fact that in physical space more localized packets have a smaller transit time through the striation.\\
All the results shown so far have been obtained starting with the arbitrary initial condition explained in  Section \ref{cp}, where the 
modes centered around $(n_0,m_0)$ have amplitudes according to the Gaussian in Eq.(\ref{gaussian}). 
Therefore, it is legitimate to wonder whether a different initial condition could result in an higher efficiency. Once again, it is not our goal to find the maximum efficiency achievable, as this problem is not very well-posed.\\
We have adopted a strategy (outlined in the Appendix) based on the fact that the linear set of Eqs. (\ref{fi2})-(\ref{chi2}) is time-reversible. Hence, one can choose an arbitrary final state, and find out what is the initial state that would produce such final state by running the simulation backwards in time.
By doing this one can see that a pre-existing population of whistler waves can enhance the efficiency of the lower hybrid to whistler conversion.\\

\section{Conclusions}\label{concl}
In this paper we have addressed the problem of linear mode conversion between lower hybrid and whistler waves through a density striation. The wave-particle interaction involving  lower hybrid/whistlers
is a key mechanism in the Earth's magnetosphere  for the loss of energetic particles in the radiation belt.
Therefore, it is important to assess the efficiency of such energy transfer.\\
We have devised a formalism based on the cold plasma linear theory which allows to study systematically the coupling between different modes, without making any assumption on the profile, width and amplitude of the striation. 
We have investigated which conditions favor an efficient conversion of energy between lower hybrid waves and whistlers. Based on the statistical findings of \citet{hoymork00} relative to the Earth's ionosphere, and consistently with the work of \citet{eliasson08}, 
the density striation has been modeled with a Gaussian profile. Although in principle these results cannot be generalized to other profiles (for instance \citet{schuck98} and \citet{tjulin04b} have studied parabolic profiles),
\citet{tjulin04b} have used both Gaussian and parabolic density profiles in their model, and concluded that the results are not strongly dependent on the profile shape. 
The conversion efficiency is generally higher for larger values of the density amplitude $\delta n$. For $\delta n=0.5$ we have found the maximum conversion efficiency to be around 50\%. Therefore, this mechanism appears to be a suitable route for energy transfer form short to long wavelength modes.\\
For a Gaussian striation, we have derived a criterion to calculate the width of the striation $D_{str}$ that would result in the optimal efficiency between the initial lower hybrid packet, and one whistler mode. 
Such criterion is $\Delta k_\perp D_{str}\approx 1.4$, where $\Delta k_\perp$ indicates the difference of perpendicular wavevectors between the resonant modes.\\
We have also presented results of the efficiency as a function of $\delta n$.\\
The conversion efficiencies shown in this paper must not be interpreted as the maximum efficiency achievable, since this would depend on the particular choice of the initial condition. 
We have suggested a strategy to increase the efficiency by suitably modifying the initial condition. For the case shown, it appears that the presence of a small quantity of whistler waves at the beginning of the simulation is beneficial to increase the conversion efficiency.\\
We have also shown that more localized packets in physical space have a smaller efficiency, suggesting that a larger transit time through the striation results in a bigger conversion efficiency.\\
Clearly the study presented in this paper does not include any kinetic effect, due to the use of the cold plasma approximation. 
Future work will address such thermal effects, by running Particle-in-Cell simulations, and therefore will result in a more complete physical description of the mode conversion process. 

\section{Appendix: time reversed-simulation}
\begin{figure}
 \includegraphics[width=20pc]{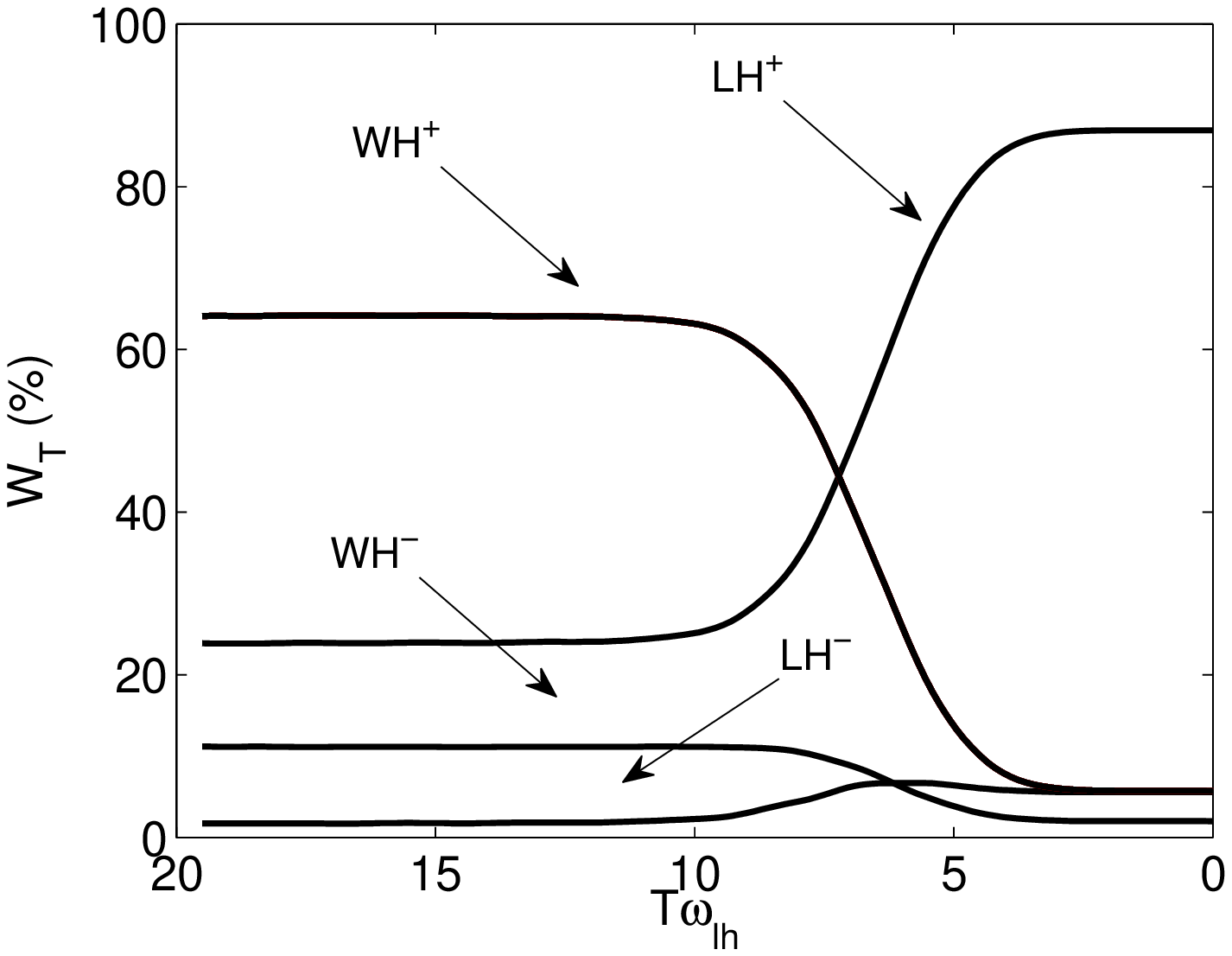}
\caption{Total energy $W_T+W_E+W_B+W_K$ in time, for the case with initial dominant mode $(n_0=26, m_0=21)$, $\delta n=0.5$, and $D_{str}=1.2$. The four curves indicate the partition of the energy among the four wave packets. The simulation is run backwards in time. The initial state is discussed in the text.}\label{fig:back}
 \end{figure}

\begin{figure}
 \includegraphics[width=20pc]{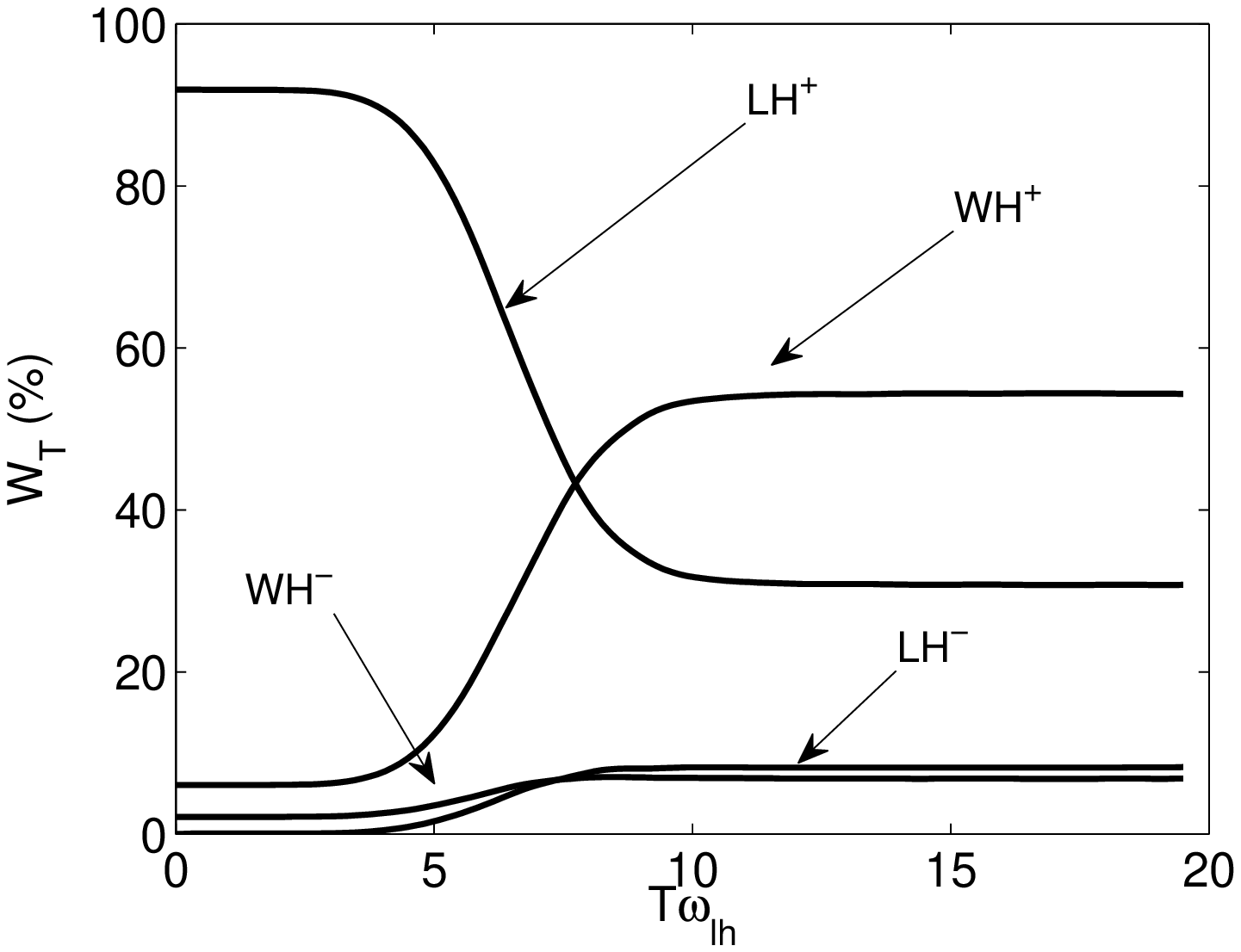}
\caption{Total energy $W_T+W_E+W_B+W_K$ in time, for the case with initial dominant mode $(n_0=26, m_0=21)$, $\delta n=0.5$, and $D_{str}=1.2$. The four curves indicate the partition of the energy among the four wave packets. The simulation is run forward in time. The initial state is discussed in the text.}\label{fig:forw}
 \end{figure}

We consider as our test bed the case with $(n_0=26, m_0=21)$, $D_{str}=1.2$, $\delta n=0.5$. From Figure \ref{fig:eff_26_21}, one can see that the efficiency of this case is about $43\%$.
The error in energy conservation is $\Delta W_T=0.3\%$. We have taken the solution at time $T\omega_{lh}=19.5$, and we have artificially doubled the amplitudes of the modes corresponding to $WH^+$ and $WH^-$. By doing this we have changed the partition of energy between the different modes, and we have created a state where $75\%$ of the energy is contained in $WH^++WH^-$, and $25\%$ is contained in $LH^++LH^-$. We have then run the simulation backwards in time. 
The time evolution of the energy content is shown in Figure \ref{fig:back}, where the time (on horizontal axis) runs from $T\omega_{lh}=19.5$ to $T\omega_{lh}=0$. The change in energy of each wave packet is the following: $LH^+$ changes from 23\% to 87\%, $LH^-$ from 2\% to 6\%, $WH^+$ from 64\% to 6\%, and $WH^-$ from 11\% to 1\%.  In principle the initial state obtained would result in an efficiency of 86\%. However, such initial state is not completely acceptable, because it is formed by a mixture of $LH^+$ and $LH^-$ modes. In other words, this is a situation where two lower hybrid wave packets, traveling in opposite direction, would hit the striation at the same time. 
For this reason we have modified this initial state obtained by the backward simulation, suppressing the $LH^-$ modes. We have then run (forward in time) the simulation with this initial condition somewhat artificially obtained. 
The energy evolution is shown in Figure \ref{fig:forw}. The changes in energy are now the following: $LH^+$ changes from 92\% to 31\%, $LH^-$ from 0\% to 8\%, $WH^+$ from 6\% to 54\%, and $WH^-$ from 2\% to 7\%. 
The total efficiency is 66\%, if computed with formula (\ref{eff_formula}).
Note that we had to choose an initial condition that already contains a small amount of whistlers. 
Therefore, in this case is probably more appropriate to take into account that the whistlers have $8\%$ of the energy at the initial time, and compute the numerator of 
formula (\ref{eff_formula}) as the gain of whistler energy (i.e. the difference between final and initial states). With such procedure the efficiency is equal to $58\%$, with a gain of $15\%$ relative to the case in Fig. 7\\
The strategy shown in Figures  \ref{fig:back} and  \ref{fig:forw} can be further improved yielding an higher conversion efficiency by starting with larger amounts of whistlers. Such study, however, 
is outside of the scope of this paper, since the initial setting would depart more and more from the setting used in the previous Section, where the initial condition is composed purely of lower hybrid modes.\\

\section{Appendix: Relationship between scalar and physical fields, in Fourier space:}

\begin{eqnarray*}
 \hat{B}_{x,nm} &=& \frac{2\pi i m}{L_y}\hat{\psi}_{nm}\\
 \hat{B}_{y,nm} &=& -\frac{2\pi i n}{L_x}\hat{\psi}_{nm}\\
 \hat{E}_{x,nm} &=& \frac{\omega_{lh}}{\omega_{ci}}\left( \frac{2\pi i m}{L_y} \frac{d\hat{\varphi}_{nm}}{dt} + \frac{2\pi i n}{L_x} \frac{d\hat{\chi}_{nm}}{dt}\right)\\
 \hat{E}_{y,nm} &=& \frac{\omega_{lh}}{\omega_{ci}}\left( \frac{2\pi i m}{L_y} \frac{d\hat{\chi}_{nm}}{dt} - \frac{2\pi i n}{L_x} \frac{d\hat{\varphi}_{nm}}{dt}\right) - \hat{V}_{z,nm} \\
 \hat{E}_{z,nm} &=& -\frac{d\hat{\psi}_{nm}}{dt}\\
 \hat{w}^i_{x,nm} &=&  \left(\widehat{\sqrt{n_{eq}}V^i_x}\right)_{nm}\\
 \hat{w}^i_{y,nm} &=&  \left(\widehat{\sqrt{n_{eq}}V^i_y}\right)_{nm}\\
 \hat{w}^e_{x,nm} &=&  \left(\widehat{\sqrt{n_{eq}}V^e_x}\right)_{nm}\\
 \hat{w}^e_{y,nm} &=&  \left(\widehat{\sqrt{n_{eq}}V^e_y}\right)_{nm}\\
 \hat{w}^e_{z,nm} &=&  \left(\widehat{\sqrt{n_{eq}}V^e_z}\right)_{nm},
\end{eqnarray*}
where 
\begin{eqnarray*}
 V^i_x (x,y)&=& \sum_m \sum_n \left(\frac{2 \pi i m}{L_y}\hat{\varphi}_{nm} + \frac{2 \pi i n}{L_x}\hat{\chi}_{nm}\right)e^{\frac{2\pi inx}{L_x}\frac{2\pi imy}{L_y}}\\
 V^i_y (x,y) &=& \sum_m \sum_n \left(\frac{2 \pi i m}{L_y}\hat{\chi}_{nm} - \frac{2 \pi i n}{L_x}\hat{\varphi}_{nm}\right)e^{\frac{2\pi inx}{L_x}\frac{2\pi imy}{L_y}}\\
 V^e_x (x,y)&=& \sum_m \sum_n \left(\frac{2 \pi i m}{L_y}\hat{\varphi}_{nm} + \frac{2 \pi i n}{L_x}\hat{\chi}_{nm} -\frac{1}{n_{eq}}\frac{\omega_{ce}}{\omega_{lh}}\frac{2\pi i m}{L_y}\hat{B}_{z,nm}\right)e^{\frac{2\pi inx}{L_x}\frac{2\pi imy}{L_y}}\\
 V^e_y (x,y)&=& \sum_m \sum_n \left(\frac{2 \pi i m}{L_y}\hat{\chi}_{nm} - \frac{2 \pi i n}{L_x}\hat{\varphi}_{nm} +\frac{1}{n_{eq}}\frac{\omega_{ce}}{\omega_{lh}}\frac{2\pi i n}{L_x}\hat{B}_{z,nm}\right)e^{\frac{2\pi inx}{L_x}\frac{2\pi imy}{L_y}}\\
 V^e_z (x,y)&=& \sum_m \sum_n \left(\hat{V}_{z,nm} - \frac{1}{n_{eq}} \frac{\omega_{ce}}{\omega_{lh}} 4 \pi^2 \left( \frac{n^2}{L_x} + \frac{m^2}{L_y}\right)\hat{\psi}_{nm}\right)e^{\frac{2\pi inx}{L_x}\frac{2\pi imy}{L_y}}
\end{eqnarray*}

%
%

\begin{acknowledgments}
We thank Dan Winske for useful discussions.
This research was conducted as part of the Dynamic Radiation Environment Assimilation Model (DREAM) project at Los Alamos National Laboratory. We are grateful to the sponsors of DREAM for financial and technical support.
This research was performed under the auspices of the
NNSA of the U.S. DOE by LANL, operated by LANS LLC
under Contract No. DE-AC52-06NA25396.
\end{acknowledgments}

\newpage


%
%
%
%
%
%
%
%



\end{article}

\newpage
%
%
%

%
%
%

%
%
%

%
%
%
%
%
%
%
%

\begin{table}
\caption{Maximum efficiency $\eta$ and error in energy conservation for several runs with initial mode $(n_0,m_0)$, $\delta n=0.5$, and $D_{str}=1$. 
The letters label the runs in Figure \ref{fig:resonance}. $\theta$ is the angle of propagation of the generated whistler mode.}\label{table1}
 \begin{tabular}{| c | c| c | c | c | c | c | c |}
 \hline
  $n_0$ & $m_0$& $k_\parallel$ &  $k_\perp$  & $\eta$ & $\Delta W_T$ & Label& $\theta$  \\ \hline
10 & 20 & 0.1575 &  1.5755 & 42.73 \% &  0.19 \% & A & $77^\circ$\\ \hline
12& 28 & 0.189 & 2.20 & 26.66 \% & 0.17 \% & B & $67^\circ$\\ \hline
14& 35 & 0.22 & 2.757 & 14.15 \% & 0.12 \% & C & $53^\circ$\\ \hline
15&17& 0.236 & 1.34 & 39.86 \% & 0.15 \% & D & $71^\circ$\\ \hline
16& 42 & 0.252 & 3.30 & 6.22 \% & 0.22 \% & E & $40^\circ$\\ \hline
17&23& 0.268 & 1.812 & 39.32 \% & 0.23 \% & F & $61^\circ$\\ \hline
18&48& 0.283 & 3.781 & 1.4\% & 0.43 \%  & G & $0^\circ$\\ \hline
19&28& 0.299 & 2.206 & 30.72 \% & 0.20 \% & H & $50^\circ$\\ \hline
21&32&0.331 & 2.521 & 23.3 \% & 0.17 \% & I & $35^\circ$\\ \hline
22&38&0.347 & 2.993 & 6.86 \% & 0.16 \% & J  & $0^\circ$\\ \hline
23&36&0.362 & 2.836 & 9.04 \% & 0.16 \% & K  & $0^\circ$\\ \hline
24&17&0.378 & 1.339 & 48.93 \% & 0.20 \% & L & $58^\circ$\\ \hline
26&31&0.410 & 2.442 & 16\% & 0.17 \%  & M  & $0^\circ$\\ \hline
26&21&0.410 & 1.654 & 45.02 \% & 0.25 \% & N & $45^\circ$\\ \hline
28&24&0.441 & 1.890 & 38.4 \% & 0.24 \% & O & $31^\circ$\\ \hline
30&27&0.473 & 2.127 & 18.45 \% & 0.18 \% & P  & $0^\circ$\\ \hline
34&23&0.536 & 1.811 & 23.16 \% & 0.19 \%  &Q & $0^\circ$\\ \hline
38&20&0.599 & 1.575 & 23.46 \% & 0.21 \%  &R & $0^\circ$\\\hline
42&17&0.662 & 1.340 & 25 \% & 0.29 \% &S& $0^\circ$\\
\hline
 \end{tabular}
\end{table}

\end{document}